\documentclass[journal]{IEEEtran}
\ifCLASSINFOpdf
\else
\fi
\usepackage{url}

\usepackage[disable]{todonotes} 
\usepackage{multirow}
\usepackage{gensymb}
\usepackage{xcolor}
\begin{document}
%
\title{Building and Evaluation of a Real Room Impulse Response Dataset}
%
%
%
\author{Igor~Sz\"{o}ke,~\IEEEmembership{Member,~IEEE,}
	    Miroslav~Sk\'{a}cel,
        Ladislav~Mo\v{s}ner,~\IEEEmembership{Student member,~IEEE,}
        Jakub~Paliesek,\\
        and Jan ``Honza'' \v{C}ernock\'{y},~\IEEEmembership{Senior member,~IEEE}
\thanks{The work was supported by Czech Ministry of Interior project No. VI20152020025 ``DRAPAK'', Google Faculty Research Award program, Czech Science Foundation under project No. GJ17-23870Y, and by Czech Ministry of Education, Youth and Sports from the National Programme of Sustainability (NPU II) project ``IT4Innovations excellence in science - LQ1602''.
We would like to thank Kamil Chalupn\'{i}\v{c}ek and Ond\v{r}ej Novotn\'{y} for helping us to collect the BUT ReverbDB.}}

%
%

\markboth{Journal of Selected Topics in Signal Processing, November~2018}%
{Szoke \MakeLowercase{\textit{et al.}}: Building and Evaluation of a Real Room Impulse Response Dataset}
%



\maketitle

\begin{abstract} 
This paper presents BUT ReverbDB - a dataset of real room impulse responses (RIR), background noises and re-transmitted speech  data. The retransmitted data includes LibriSpeech test-clean, 2000 HUB5 English evaluation and part of 2010 NIST Speaker Recognition Evaluation datasets. We  provide a detailed description of RIR collection (hardware, software, post-processing) that can serve as a ``cook-book'' for similar efforts. We also validate BUT ReverbDB  in two sets of automatic speech recognition (ASR) experiments and draw conclusions for augmenting ASR training data  with real and artificially generated RIRs. We show that a limited number of real RIRs, carefully selected to match the target environment, provide results comparable to a large number of artificially generated RIRs, and that both sets can be combined to achieve the best ASR results. The dataset is distributed for free under a non-restrictive license and it currently contains data from 8 rooms, which is growing. The distribution package also contains a Kaldi-based recipe for augmenting publicly available AMI close-talk meeting data and test the results on an AMI single distant microphone set, allowing it to reproduce our experiments. 
\end{abstract}

\begin{IEEEkeywords}
far-field, automatic speech recognition, room impulse response, reverberation, SineSweep, Maximum Length Sequence, noise, deep neural network, Kaldi, AMI.
\end{IEEEkeywords}

%
\IEEEpeerreviewmaketitle

%
%
%
%

\section{Introduction}
\IEEEPARstart{A}{utomatic} speech recognition  (ASR) has made tremendous improvements in the last decade and services  and applications making use of close-talk speech (such as SMS dictation, personal assistants, or contact-center speech data analytics) are on the market and serving millions of customers. On the other hand, ASR  from far-field microphones is far less advanced and significant research efforts are devoted to improving its performance and robustness. 

Despite all the research efforts, the best one can do to obtain a decent ASR performance is to collect transcribed data from the target domain. For far-field ASR, however, this is unfeasible due to the infinity of different room configurations, microphone placements, microphone types, noise conditions, etc. 
\emph{Data augmentation} --- reverberation of source  data  using estimated or artificially generated room impulse responses (RIR) and adding real noises to simulate the environment ---  is therefore the most common technique to build a robust ASR~\cite{karafiat2017} nowadays.  

Collecting noises is easy as there are lots of public sources and the noises can also be extracted from existing speech data. On the other hand, gathering  real RIRs is technically difficult and time demanding. To overcome this problem, artificial RIRs are usually used as they can be  generated automatically and in large quantities. They are good enough in scenarios, where the loudspeaker and microphone are facing each other~\cite{Melot2015} (see Section~\ref{sec_mic_occlusion}), but simulating RIRs for microphones, partly or fully hidden, is not widely supported by existing tools. Here, the estimation of real impulse responses is the only way.

There is also a lack of ``parallel audio corpora'' where both clean close-talk (ideally anechoic) speech is available together with  reverberated and noised version in various environments. This parallel corpus may also be useful in scenarios such as audio enhancement, denoising, dereverberation or beam-forming.

\subsection{Motivation and goals}

The motivation of this paper is to: a) introduce the Brno University of Technology Speech@FIT Reverberation Database (\textbf{BUT ReverbDB}) and describe the methodology of its collection; and b) compare the impact of data augmentation using either artificial or real impulse responses in scenarios with no target training data available for the development of an ASR system. BUT ReverbDB contains also data for developing and testing of far-field  Speaker REcognition (SRE) system~\cite{Mosner2018a,Mosner2018b}, but this paper concentrates solely on ASR.

The BUT ReverbDB was built in order to collect a large number of various RIRs, room environmental noises (or ``silences''), retransmitted speech (for ASR and SRE testing), and meta-data (positions of microphones, loudspeakers etc.). The goal is to provide the speech community with a dataset for data augmentation and distant microphone or microphone array experiments in ASR and SRE. The database is distributed under Creative Common 4.0 Attribution license (CC-BY 4.0 -- free for commercial, academic, and government use) and is available on the BUT web pages\footnote{\url{https://speech.fit.vutbr.cz/software/but-speech-fit-reverb-database}}. 

So far, BUT ReverbDB contains data from  $8$ rooms (large, middle and small size). We placed $31$ microphones in each room. The loudspeaker was usually placed at $5$ different positions per room. We measured room impulse responses, environmental noise (silence) and we retransmitted \textbf{Libri-Speech Test-clean} dataset~\cite{Panayotov2015}, \textbf{2000 HUB5 English evaluation} set\footnote{\url{https://catalog.ldc.upenn.edu/LDC2002S09}} and also part of \textbf{NIST Speaker Recognition Evaluation 2010} dataset~\cite{Greenberg2010} (the availability of the HUB5 and SRE data is limited to sites that have a valid LDC license to the original data).

All loudspeaker and microphone positions are measured and stored in meta-files in  Cartesian and polar coordinates, and in an absolute and relative (to the loudspeaker) way.

BUT is taking part in the ``DRAPAK'' project sponsored by the Czech Ministry of Interior concentrating on ASR and SRE in the security domain (including  close-talk and distant microphones, listening devices, etc.), therefore, the motivation of BUT ReverbDB is to collect acoustic environments which are  challenging and cannot be easily simulated. That is also why our microphones are partly placed in very unusual places.

A number of ASR experiments were performed with BUT ReverbDB --- partly as a sanity check and partly to show the importance of real environment impulse responses and background noises for training data augmentation. 

\subsection{Organization of the paper}
The paper is organized as follows: the following Section~\ref{sec_relatedwork} presents related work in a distant microphone ASR,  existing RIR data-sets and their shortcomings. Section~\ref{sec_howtogetRIR} summarizes approaches in estimating real and computing simulated RIRs. Section~\ref{sec_reverbdb} presents BUT ReverbDB with details and practical recommendation in Appendix~\ref{sec_ReverbDB_appendix}. Section~\ref{sec_czechexp} describes the first ASR experiments aimed to  validate the data-set.  Section~\ref{sec_amiexp} presents training data augmentation work on AMI data -- these experiments are fully reproducible as all RIRs, data and recipes are made available.  Section~\ref{sec_conclusion} concludes the paper and outlines future work. 

\section{Related work}\label{sec_relatedwork}
\subsection{Automatic speech recognition on reverberated data}

ASR performance heavily degrades when facing a mismatch between training and evaluation data conditions~\cite{Yoshioka2012}. Such a mismatch  can include   the environment (background  noise,  recording  conditions  (microphones and rooms)) and  speaker characteristics (calm speech versus shouting with Lombard effect).
An obvious solution is to collect and transcribe data from the target domain. However,  when ASR is used in the field, the time, effort, and cost of transcribing data for the new conditions becomes limited (as in IARPA's BABEL\footnote{https://www.iarpa.gov/index.php/research-programs/babel} and DARPA's LORELEI\footnote{https://www.darpa.mil/program/low-resource-languages-for-emergent-incidents} projects) or prohibitive (ASpIRE challenge~\cite{Harper2015}). 

Changes in room acoustics can be a significant source of mismatch (and hence an ASR word accuracy drop) as was shown in the International  Computer  Science  Institute  (ICSI)  meeting  room dataset~\cite{Janin2003,Parthasarathi2013}, Augmented Multi-party Interaction (AMI) meeting room corpus~\cite{Hain2007}, and the Multi-channel Wall  Street  Journal  Audio  Visual  Corpus  (MC-WSJ-AV)  corpus~\cite{Lincoln2005}.

The ASpIRE challenge~\cite{Harper2015} addressed far-field microphone recordings of conversational speech with a very large vocabulary. The test data differed substantially from the training and development data. The ASpIRE challenge demonstrated that working continually on  the  same  test  data  and  making  progress  on  that  data  may  not  guarantee   robustness   to   data   collected   in   new (although   related)   recording conditions.  Reverberation was clearly important in both the   development   and   evaluation   sets;   however,   
microphone variability was greater in development set (Mixer $6$~\cite{BRANDSCHAIN2010}) and room variability in the evaluation set (Mixer $8$).
This  suggests  that  new  challenges  that  aim  to  measure  system  robustness  need  to  creatively  collect  new  test  data  with  mismatch  and   then   limit   testing   on   these   data   until   after   systems   are   developed. 

An interesting analysis of ASpIRE  results~\cite{Melot2015} studied the correlation of source-to-microphone distance and ASR performance, and concluded that rather than trying to extrapolate ASR performance from simple distance metrics, one needs to also take  into account the orientation of both the speaker and the microphone. This means that we do need not only data with microphones facing directly the speaker, but also  other, more complicated, speaker-microphone positions.  

Another paper by Ko~\cite{Ko2017} based on  ASpIRE and  AMI data pointed out that the performance gap between using simulated and real RIRs can be eliminated when point-source noises are added. For Ko, the trained acoustic model not only performed well in the far-field scenario, but also provided better results in the close-talking one.

The problem of robustness of ASR on distant microphones was also  approached by a series of CHiME challenges. CHiME-1~\cite{Barker2013} aimed at small vocabulary ASR (command and control) in a real  living room using binaural microphones. Target speech commands were  mixed into the environment noises at a fixed position using genuine room impulse responses.  CHiME-2~\cite{Vincent2013} used the CHiME-1 dataset and aimed at a larger vocabulary  and a more realistic mixing process accounting for small head movements  while speaking. 
CHiME-3~\cite{Barker2015} and CHiME-4 are designed around the popular Wall Street Journal (WSJ) corpus and feature talkers speaking in challenging noisy environments recorded using a $6$-channel tablet-based microphone array. They consist of two types of data 1) ``Real data'' -- read speech data  recorded in real noisy environments (on a bus, cafe, pedestrian area, and street junction) uttered by actual talkers; and 2) ``Simulated data'' -- noisy utterances  generated by artificially mixing clean speech data with noisy backgrounds.
Actually, CHiME-5~\cite{Barker2018} aims to be the first large-scale corpus of real multi-speaker conversational speech recorded via commercially available multi-microphone hardware (Kinect and binaural microphones) in multiple homes. Speech material was gathered from  a 4-people dinner party scenario in 20 homes. However,  no RIRs were collected in any CHiME data collections. 

In REVERB challenge~\cite{Kinoshita2013}, the goal was to evaluate different approaches to ASR and speech enhancement  on  simulated data (WSJ artificially reverberated and noised by real world RIRs and noises) and real data (WSJ utterances read by humans in real noisy and reverberant conditions).
The conclusion of the REVERB challenge~\cite{Kinoshita2016} was ``Apart from the problems of ASR techniques,
concerning the data preparation stage, challenges remain in simulating acoustic data that are
close to actual recordings. Developing better simulation techniques remains another important research
direction since simulations can be useful to evaluate techniques and generate relevant training data for
acoustic model training.''

The results of Ravanelli~\cite{Ravanelli2017} show that using real RIRs to augment the training data provides a significant improvement on the ASR Word Error Rate (WER) (using a recent deep neural network system) to the data augmentation using just artificial RIR even with setting the room parameters as close as possible to the real room.

\subsection{Available room impulse responses sets}

\begin{table*}[htb]
\centering
\begin{tabular}{|l|c|c|c|c|c|c|c|}
\hline
\multicolumn{1}{|c|}{Name}         &   \# RIRs   & \# Rooms  & $RT_{60}$     & M2L dist. [m]   & Metadata     &  Target       & IR type     \\ \hline
ACE          &    $700$    & $7$      & $0.34-1.25$    & $0.5-4.0$       & very good    &  DRR and $RT_{60}$ evals       & ESS \\ \hline
AIR          &    $214$    & $6$      & $0.12-0.78$    & $0.5-10.0$      & good         &  SE, binaural             & MLS  \\ \hline
REVERB       &    $24$     & $3$      & $0.25-0.70$    & $0.5-2$         & N/A           &  SE, ASR                  & N/A      \\ \hline
RWCP         &    $3k$     & $9$      & $0.00-1.30$    & $2-4$*          & N/A           &  SE, ASR                  & TSP \\ \hline
\hline
BUT ReverbDB &    $1.3k$   & $8$      & $0.59-1.85$    & $0.5-15.0$      & excellent    &  SE, ASR                  & ESS (MLS)\\ \hline
\end{tabular}
\vspace*{3mm}
\caption{Comparison of publicly available RIR datasets. M2L means Microphone to Loudspeaker distance -- * denotes our guess from photos. Metadata means recording protocols including information as photos, placing coordinates, type of microphones, room dimensions and equipment -- N/A denotes Not Available.} 
\vspace*{-5mm}
\label{tab_RIR_datasets}
\end{table*}

In the past, several attempts of collection of RIRs and environmental noises were done, either for research purposes in the field of speech enhancement, speech recognition, beam-forming, acoustic environment characterization, or for smart-homes. We identified two main categories of datasets:

\noindent 1) Designed for Speech Enhancement (SE) and Automatic Speech Recognition (ASR) (see Table~\ref{tab_RIR_datasets} for details):
  \begin{itemize}
    \item Aachen Impulse Response (\textbf{AIR})\footnote{\url{https://www.iks.rwth-aachen.de/fileadmin/user_upload/downloads/forschung/tools-downloads/air_database_release_1_4.zip}}~\cite{Jeub2009} database ($6$~types of room, with several configurations of microphone/source placing, including the binaural microphone) aims at evaluation of speech enhancement algorithms dealing with room reverberation. 
    \item \textbf{ACE} Corpus\footnote{https://acecorpus.ee.ic.ac.uk/}~\cite{Eaton2015e} ($50$ microphones in $6$ devices, placed in $2$ setups in $7$ rooms) was used in ACE challenge~\cite{Eaton2016} of $T_{60}$ and Direct-to-Reverberant Ratio estimation methods using real noisy reverberant speech. The recording devices are a mobile phone, notebook and $32$-channel spherical microphone array.
    \item \textbf{REVERB} challenge\footnote{\url{https://catalog.ldc.upenn.edu/LDC95S24}}~\cite{Kinoshita2013} dataset ($2$ times $3$ types of room, near and distant microphone placement, $2$ microphone angles) is a common evaluation framework including datasets, tasks, and evaluation metrics for both speech enhancement and ASR. It is carefully designed to assess robustness against reverberation. It contains WSJCAM0~\cite{Lincoln2005} utterances, either spoken by humans in reverberant conditions or artificially retransmitted by a loudspeaker.
    \item \textbf{RWCP} Sound Scene Database\footnote{~\url{http://research.nii.ac.jp/src/en/RWCP-SSD.html}}~\cite{Nakamura2000} (circular and linear microphone array placed in $9$ rooms with several positions of the loudspeaker) is a data collection project that serves sound source localization, retrieval, recognition and speech recognition in real acoustical environments. It includes retransmitted phonetically balanced sentences with precise position tracking of moving loudspeaker.
  \end{itemize}

\noindent 2) Designed for smart-home appliances:
  \begin{itemize}
    \item \textbf{DIRHA} project\footnote{\url{http://dirha.fbk.eu/English-PHdev}}~\cite{Ravanelli2015} dataset is composed of real phonetically-rich sentences recorded in a domestic environment equipped with a large number of microphones and microphone arrays distributed in space. It has very precious material for studies on multi-microphone speech processing and distant-speech recognition. No RIRs are public.
    \item \textbf{VoiceHome}~\cite{Bertin2016} corpus aims at command and control, and dialog scenarios (smart-home). Reverberated and noisy speech spoken by $12$ native French talkers in $4$ houses ($3$ rooms per house) is recorded by an $8$-microphone device at various angles and distances and in various noise conditions. $188$ RIRs were collected, however, none are publicly available.
    \item \textbf{Sweet-Home}~\cite{vacher2014} corpus also targets command and control scenario (smart home). It consists of $26$ hours of speech data (French) recorded in $4$ rooms ($1$ flat), $7$~channels (2+2+2+1). No RIRs were recorded.
    \end{itemize}

The \emph{SE/ASR datasets} are mainly focused on RIR estimation and ambient noise collection. 
They expect the microphones to be integrated in devices placed on furniture and at a reasonable distance from the loudspeaker (up to $4$ meters). The \emph{smart-home datasets} are focused more on the command and control scenario in reverberant and noisy environments. They contain recorded sets of proprietary utterances spoken by several humans (VoiceHome and Sweet-Home in French). Microphones are expected to be integrated in walls/ceiling, as small microphone arrays. A drawback is the lack of RIRs unavailability, although they were estimated. 

Overall, none of these datasets include retransmitted publicly available speech data. Next, a majority of the datasets contain several microphone arrays which limits the variability in microphone positions. An interesting point was raised by Ravanelli~\cite{Ravanelli2014} who found that for the ASR adaptation, variability across rooms is more important than within the room. So from our opinion it does not make much sense to place large microphone arrays in few rooms. Lastly, all datasets expect ``cooperating user'' by placing microphones on a furniture (smart assistants, handheld devices, etc) at a reasonable distance or integrated in walls / ceiling using small arrays with direct human to microphone visibility. In our opinion, there is a clear lack of:
\begin{enumerate}
  \item ad-hoc microphone placement in ``non-cooperative'' positions (large obstacles, partly or fully hidden microphone) where the user is not even aware of the presence of a microphone. 
  \item retransmissions of publicly available data. Here we note that according to our experiments, this type of data can be artificially generated by reverberating the source data and adding particular noise (see Section~\ref{sec_czechexp}).
  \item good metadata as many datasets contain RIRs without precise microphone / loudspeaker placing and orientation coordinates and other description. This is fine for ``put all data on one heap and train a DNN'' scenario, but it is not sufficient for any deeper analysis. Precise metadata may also be used for experiments comparing real and artificial RIRs (see Section~\ref{sec_howtogetRIR}).
  \item variety in acoustic environments. All available public data-sets together contain RIRs from only $25$ rooms (mainly offices, meeting and lecture rooms).
\end{enumerate}

A good RIR dataset for SE/ASR should have a good variety over environments (rooms), microphone placing (visible, hidden), microphone types (high-end, MEMs, low-end, handheld device, integrated etc.), and precise metadata. We do not consider microphone arrays as important since there can be many variations. However, the single distant microphone still has a significant application coverage.

In conclusion, a large data set of RIRs with consistent recording protocols covering standard acoustic environments like offices, houses, corridors, cars etc., is missing. The closest RIR datasets are the ACE and AIR.  Our goal --- as our target application is speech data mining (ASR and SRE) from a variety of sources (table top microphones, IoTs, mobile devices, smart assistants, smart homes, but also listening devices, bugs and other non-standard microphones) --- is to have RIRs from a variety of microphone positions.

\section{Obtaining room impulse responses}\label{sec_howtogetRIR}
An RIR can be obtained in two principal ways: the first is to measure the environment and obtain  the ``real'' RIR,  the second is to generate it artificially by a simulation. 

\subsection{Real room impulse responses}
Several methods were developed to measure the real RIR. The Maximum Length Sequence (MLS) technique was first proposed by Schroeder~\cite{Schroeder:1979}. Other techniques were suggested to reduce distortion artifacts of MLS such as the Inverse Repeated Sequence (IRS)~\cite{Dunn1993}. Another method -- Time-Stretched Pulses -- was proposed by Aoshima~\cite{Aoshima1981}. Finally, a logarithmic Sine Sweep technique introduced by Farina~\cite{Farina2000} should overcome some limitation of the other ones. 

We briefly summarize these techniques and refer the reader to Stan et al.~\cite{Stan:2002} for extensive comparison with a supporting mathematical apparatus:

{\bf Maximum Length Sequence} is based on the excitation of the acoustical space by a periodic pseudo-random signal~\cite{Golomb1981}. The number of samples of one period of MLS signal is: $L=2^{m}-1$, where $m$ is the order.
The RIR is then calculated by circular cross-correlation between the measured output and the original MLS signal. The circular cross-correlation obviously causes a well known problem~\cite{Stan:2002} -- the time-aliasing error, which can be overcome by setting $L$ longer than expected RIR measured (considering $48kHz$ sampling frequency, the $m > 17$ to be on the safe side). The MLS method has a strong immunity to signals not correlated with the excitation signal, due to the MLS phase spectrum being irregular and a uniform density of probability. Any disturbing signals are ``spread'' uniformly along the deconvolved RIR. Using averaging as post-processing leads to the reduction of the distortions. This makes the MLS suitable for RIR measuring in an occupied room or exterior setting.
On the other hand, a major drawback is in the appearance of ``distortion peaks''~\cite{Vanderkooy2012}. The MLS method relies on the assumption of Linear, Time-Invariant (LTI) system. Any inherent non-linearities of the measurement system (especially the loudspeaker) are present in the RIR and appear as cracking sounds when convolved with an audio. They can be partly avoided by precise calibration (mainly the loudspeaker output level). MLS also expects input/output sampling clock synchronization~\cite{Farina2000}.

{\bf Inverse Repeated Sequence} reduces the ``distortion peaks'' drawback of MLS. The IRS excitation signal is a sequence of length $2L$, the first half is equal to MLS and the second half is inverse MLS~\cite{Dunn1993}. The rest is common with the MLS method (circular cross-correlation, input/output sampling clock synchronization, immunity to disturbing signals). 

{\bf Time-Stretched Pulses} method reduces the distortion peaks by the expansion and compression of an impulsive signal~\cite{Aoshima1981}. 
It also relies on the assumption of LTI system. According to the spectral properties of stretched pulses, this method is not immune to disturbing signals (it cannot be used in occupied rooms~\cite{Stan:2002}).

{\bf Exponencial Sine Sweep -- (ESS)} uses an exponential time-growing frequency sweep as the excitation signal. ESS does not rely on the LTI system assumption in contrast to the MLS, IRS, and TSP. It is possible to perform simultaneous deconvolution of the linear impulse response of the system and selective separation of each impulse response corresponding to the harmonic distortion using the excitation signal. The harmonic distortions appear prior to the linear impulse response~\cite{Farina2000}.
The impulse response deconvolution process is implemented by the linear convolution of the measured output with the analytical inverse filter estimated from the excitation signal. The advantage upon MLS and IRS methods is that linear convolution overcomes time-aliasing problems. If the emitted ESS is shorter than the RIR to be measured, we just need sufficient silence to be added at the end of the ESS to recover the tail of RIR.
The ESS method is perfect in rejecting the harmonic distortions as they appear prior to the ``linear'' impulse response estimation. It has an excellent RIR signal-to-noise ratio. It also does not need an output level calibration. On the other hand it is not immune to disturbing signals and is suitable for quiet rooms~\cite{Stan:2002}.

From the experimental point of view, according to~\cite{Ravanelli2012}, the Exponential Sine Sweep (ESS) has shown robustness to  changing  loudspeaker output level while MLS and LSS (Linear Sine Sweep) tend to degrade the ASR  WER in the presence of  higher output volumes.
ESS was also found  robust (only $0.5\%$ WER deterioration) when switching from expensive studio monitors to cheap PC loudspeakers.

In conclusion, the best method for our needs is the Exponential Sine Sweep~\cite{Farina2007} as it is not sensitive to  output level calibration and we will use it in empty environments (non-occupied rooms). We accompany the ESS measurements with MLS to have the RIR in cases, where a microphone is placed close to a noise source and the SNR is low for this particular microphone. Our hardware setup (see Appendix~\ref{sec_ReverbDB_appendix}) also does not have clock signal synchronization  between playback and recording device which limits the use of MLS. However, we were able to compensate this by re-sampling the recorded MLS signal (see the following section). We use MLS implementation by Thomas~\cite{Thomas2009} and ESS implementation available as a free Matlab code\footnote{\url{http://freesourcecode.net/matlabprojects/69639/exp.sweep-and-impulse-response-in-matlab}}.

\subsection{MLS - Compensation of clock asynchronicity}\label{sec_MLSsynccomp}

\begin{figure}[!t]
\centering
\includegraphics[width=8cm]{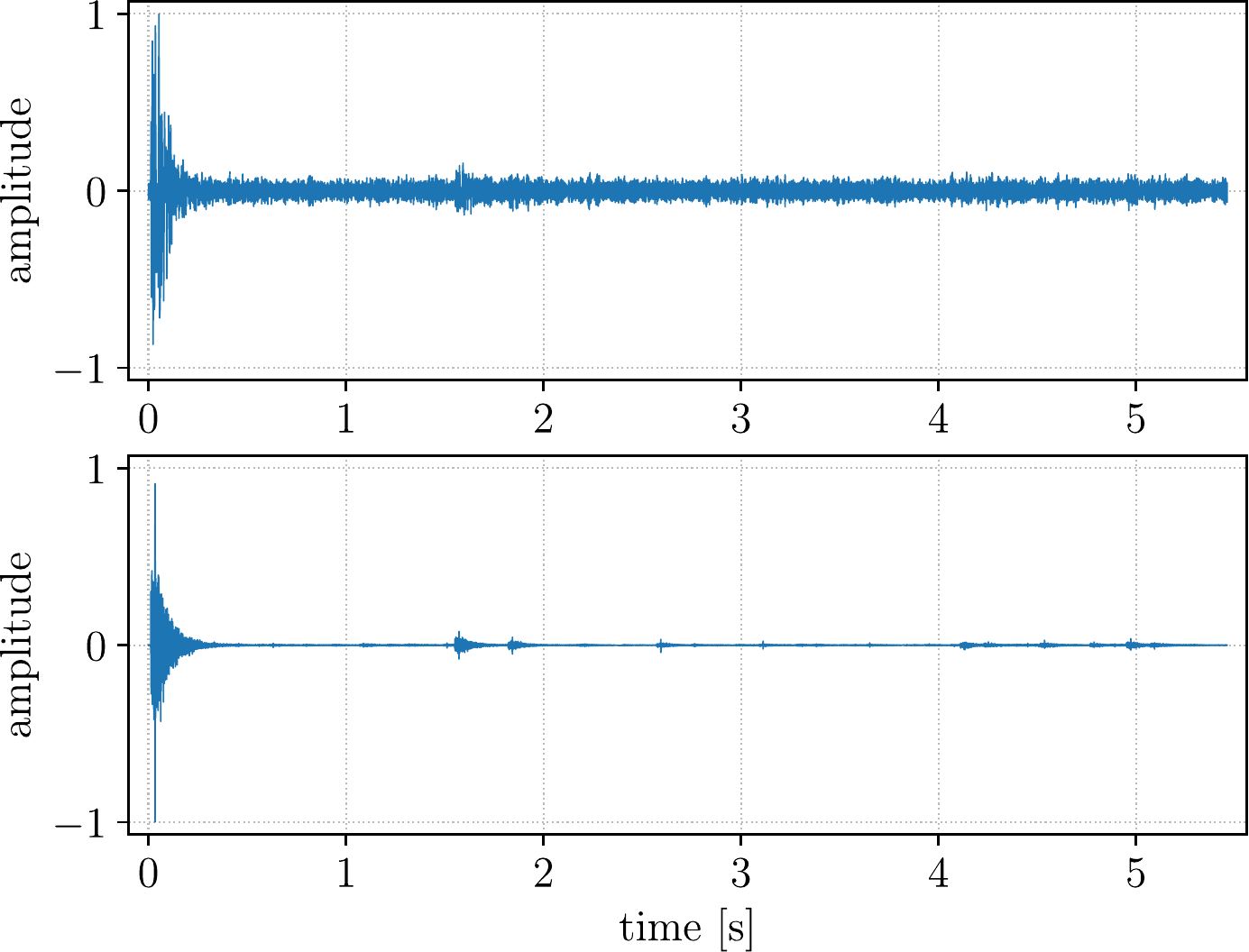}
\caption{Top panel shows a RIR estimated by MLS  without playback and recording device clock synchronization. Notice the noise in the late reflections. Bottom panel shows RIR when the recorded MLS sequence was re-sampled to match the playback sampling frequency.}
\label{fig_mlssync_mlsplot}
\end{figure}

The playback / recording clock asynchronicity causes a time stretch of a recorded signal compared to the excitation one. It lead to distortions of measured RIR when applying the circular cross-correlation on the stretched signal. We conducted an experiment where we compensated the difference in clocks for the playback and recording device. We applied the cross-correlation function on the first and last recorded period of the MLS signal (we use $32$ repetitions of the MLS sequence of order $m=18$).
The time shift was then applied in the re-sampling of the recorded MLS sequence in order to match the played one sample-to-sample (see Figure~\ref{fig_mlssync_mlsplot} for RIR with and without the sampling frequency compensation).

Finally, we did an ASR experiment (see Section~\ref{sec_czechexp} for more details) where RIRs of two rooms were estimated for $31$~microphones. The test data was then artificially reverberated and processed by the ASR, and we compared word accuracies of MLS- and ESS-processed test data. The average difference between compensated MLS and ESS is only $0.37\%$ absolute on word accuracy. This shows that the compensated MLS method provides very similar RIRs to the ESS method.

Anyway, we decided not to use MLS in further experiments and stuck to ESS, but we continued recording both MLS and ESS signals and let the user choose BUT ReverbDB.

\subsection{Artificial room impulse responses}\label{sec_artir_rirest}
For the purpose of an artificial RIR generation, computer simulation must be performed. Approaches that have been developed may be roughly divided into two groups: wave-based and ray-based methods~\cite{Siltanen2010,Savioja2015}. The former techniques are designed to solve the wave equation, whereas the latter group makes use of geometrical acoustics where sound propagates in form of rays and wave nature is neglected. Wave-based methods provide more realistic results since they are inherently able to simulate sound propagation phenomena such as diffraction. However, this advantage comes at the cost of computational expense. The boundary element method~\cite{Ciskowski1991} and finite element method~\cite{Ihlenburgc1998}, representatives of the wave-based group, discretize surface or volume to elements that interact according to the wave equation which is costly. This is a limitation because when augmenting training speech data, numerous different room conditions must be simulated.

Therefore, ray-based methods are more suitable for our purpose. Image Source Method (ISM) formulated by Allen~\cite{Allen1979} and ray tracing~\cite{Krokstad1968} are well-known techniques based on geometrical acoustics. In ray tracing, a sound source generates multiple rays that are cast to a room at a single time instance. They propagate through free space and get reflected on walls and obstacles. Each reflection decreases ray energy according to the absorption of the material. RIR is then created using rays that passed through a receiver and their energies.

The ISM  uses  ``unwrapping'' of room geometry. Every reflection of the sound ray from a wall can be considered as a sound ray originating from a virtual source behind the wall. The sound ray energy is reduced by the wall reflection coefficient (absorption). Using this principle, the room geometry is unfolded several hundred or thousand times and appropriate virtual sound sources are placed in the space. The final RIR is a summation of delayed Dirac impulses passed through a low-pass filter (to respect the sampling theorem) and attenuated by an appropriate number of ``walls'' it has to reflect from.

In the speech community, the ISM is prevalent when it comes to data augmentation~\cite{Kim2017} and multiple toolkits have been created~\cite{Habets2010,Scheibler2018}. To the best of our knowledge, there is no extensive study comparing ray tracing and image source method for data augmentation and showing the superiority of ray-tracing. For this reason, we use the artificial RIR generator implemented by Habets~\cite{Habets2010}.  It allows for setting reflection coefficients of particular walls and orientation and directional characteristics of microphones. An omnidirectional loudspeaker is considered in the simulation.

\section{BUT ReverbDB}
\label{sec_reverbdb}

So far, we measured $8$ rooms with the majority of data processed, exported and made available. The available rooms are summarized in Table~\ref{tab_rooms_measuredsofar}. The volume is an approximation for non-block shape rooms. The number of RIRs is given by the number of microphones times number of loudspeaker positions. The number of retransmissions (column ``Ret.'') indicates how many times the speech data (LibriSpeech Test-clean, $2000$ HUB$5$ English evaluation set, and NIST SRE $2010$) was retransmitted. While RIR data was recorded for each loudspeaker position, the audio was not retransmitted for all of them, as it is a very time consuming process.

\begin{table}[tb]
\setlength{\tabcolsep}{3.2pt}
\centering
\begin{tabular}{|l|c|c|c|c|c|c|}
\hline
    \multicolumn{1}{|c|}{Room}  & \multicolumn{1}{c|}{Dimensions} & \multicolumn{1}{c|}{Volume} & \multicolumn{1}{c|}{$RT_{30}$} & \multicolumn{1}{c|}{RIRs} & Ret. & Type       \\
    \multicolumn{1}{|c|}{ID}  & \multicolumn{1}{c|}{[m $\times$ m $\times$ m]}     & \multicolumn{1}{c|}{[m$^3$]} & \multicolumn{1}{c|}{[s]} & \multicolumn{1}{c|}{[\#]} & & \\ \hline
\textbf{\emph{Q301}}  & 10.7 $\times$ 6.9 $\times$ 2.6   & 192    &  0.78 & $31 \times 3$  & 1      & office     \\ \hline
\textbf{\emph{L207}}  & 4.6 $\times$ 6.9 $\times$ 3.1    & 98     &  0.61 & $31 \times 6$  & 2      & office     \\ \hline
\textbf{L212}         & 7.5 $\times$ 4.6 $\times$ 3.1    & 107    &  0.70 & $31 \times 5$  & 2      & office     \\ \hline
\multicolumn{1}{|c|}{\textbf{R112}}         & \multicolumn{1}{c|}{4.4 $\times$ 2.8 $\times$ 2.6*}   & \multicolumn{1}{c|}{{\scriptsize $\sim$}40}   &  \multicolumn{1}{c|}{0.59} & \multicolumn{1}{c|}{$31 \times 5$}  & \multicolumn{1}{c|}{0}      & \multicolumn{1}{c|}{hotel room} \\ 
         & \multicolumn{1}{c|}{2.2 $\times$ 1.2 $\times$ 2.6*}   &    &   &  &  &  \\ 

\hline
L227                  & 6.2 $\times$ 2.6 $\times$ 14.2   & 229    &  1.85 & \,$31 \times 11$ & 3      & stairs     \\ \hline
CR2                   & 28.2 $\times$ 11.1 $\times$ 3.3  & 1033   &  1.59 & $31 \times 4$  & 0      & conf. room \\ \hline
E112                  & 11.5 $\times$ 20.1 $\times$ 4.8* & {\scriptsize $\sim$}900  &  1.17 & $31 \times 2$  & 0      & lect. room \\ \hline
D105                  & 17.2 $\times$ 22.8 $\times$ 6.9* & {\scriptsize $\sim$}2000 &  1.13 & $31 \times 5$  & 1      & lect. room  \\ \hline
\end{tabular}
\vspace*{3mm}
\caption{List of rooms in the current distribution of BUT ReverbDB. The star denotes rooms with non-block shape (for example an ``L'' shape). The room volume is an approximation. The number of RIRs consists of the number of microphones times number of loudspeaker positions. Column ``Ret.'' indicates  number of speech data retransmissions. Rooms used in the test data experiments (Section~\ref{sec_czechexp}) are noted in \textit{italics}, $4$ rooms used in the training data augmentation experiments (Section~\ref{sec_amiexp}) are noted in \textbf {bold}.} 
\vspace*{-5mm}
\label{tab_rooms_measuredsofar}
\end{table}

We plan to continue in the collection of BUT ReverbDB.  Our goal is about $50$ in-door environments including cars. We also plan to increase the number of devices by using a $2^{nd}$ order ambisonic microphone, MEMS microphones, tablets, mobile phones and headsets (see Appendix~\ref{sec_ReverbDB_appendix} for more technical details).

\section{ASR test data experiments}\label{sec_czechexp}

In this section, we describe experiments conducted on ASR test data. Initially, this work was intended as just a set of BUT ReverbDB sanity checks, but we found that several topics are of general interest.

To begin with, we show that we are able to artificially retransmit (convolving with a RIR) test data and obtain the same word accuracy (WAC) as with the real retransmitted data. This leads to the conclusion that retransmission of acoustic data can be substituted with RIR estimation and noise recording, requiring much less time spent in the physical room.

We have also verified the influence of background noise on data augmentation reported in~\cite{karafiat2017,Ko2017} and confirmed that adding noise is helpful.

The influence of microphone occlusion on RIR estimation was investigated too. Theoretically, the RIR of occluded microphone can be sythesized, however, we have not yet found any tool ready to use it (see Section~\ref{sec_howtogetRIR}). We have shown that while the ISM method is good enough for non-occluded microphone placing, when the microphone is hidden, the real RIR is a clearly superior method. This further supports the need of real RIR measuring.

\begin{table}[htb]
\centering
\begin{tabular}{|l|c|c|c|}
\hline
\multicolumn{1}{|c|}{Data}                & Total            & Test-set & Type

\\ \hline
\multirow{3}{*}{SpeeCon~\cite{Glembek2006}} & \multirow{3}{*}{$759.4$h ($+996.2$h)}  &  \multirow{3}{*}{$69.9$m / $15$} & prompted, \\
& & & close talk,\\
& & & distant mic.
\\ \hline
\multirow{4}{*}{Third party}         & \multirow{4}{*}{$641.7$h ($+1128.8$h)}  & \multirow{4}{*}{$22.9$m / $14$} & prompted, \\
& & & spontaneous, \\
& & & close talk, \\
& & & distant mic.
\\ \hline
\multirow{2}{*}{Ministry of Def.} & \multirow{2}{*}{$140.0$h ($+247.3$h)}  & \multirow{2}{*}{none} & spontaneous,\\
& & & telephone
\\ \hline
\hline
SUM                 & $3913.4$h           & $92$m / $39$ & - \\ \hline
\end{tabular}
\vspace*{3mm}
\caption{Data sources used for the test data experiments. Augmented data amounts are in brackets. We used a mix of reverberation using  RIRs  generated by ISM and additive noises. "Test-set" denotes duration (in  minutes) and the number of speakers used for ASR experiments in this paper.}
\vspace*{-5mm}
\label{tab_cze_trainingdata}
\end{table}

We used a pre-trained Czech ASR based on stacked-bottleneck architecture~\cite{Grezl2014}. The $8$kHz training data consists of $3900$hrs of telephone speech, close talk data, distant microphone data and augmented data (RIRs artificially generated by ISM and a set of publicly available noises\footnote{\url{http://freesound.org}}). See Table~\ref{tab_cze_trainingdata} for further details. The vocabulary and language model were derived from acoustic data transcriptions. We considered this recognizer as robust enough to provide us meaningful results. We adapted neither the acoustic model nor the language model on the test data (no speaker adaptation, no NN fine-tuning, etc.). All results are reported as word accuracy (WAC).

We selected a reasonable test-set to conduct experiments and retransmitted it in various environments. We used only clean close-talk data without reverberation and noise in the background as a source for retransmission: $92$ minutes of prompted speech and phonetically balanced sentences from $39$ speakers (gender and age balanced) -- see Table~\ref{tab_cze_trainingdata}. We achieved $75.9\%$ in word accuracy on the clean test-set; this is our \textbf{baseline}. We used the reference speech/non-speech segmentation in decoding the retransmitted data in further experiments, in order to suppress  the influence of Voice Activity Detection (VAD) on  overall results and conclusions.

We denote \textbf{Retransmit} (real retransmission) the test-set which was replayed in the particular room $r$ and hence recorded with the room's natural reverberation and background noise by microphone $c$. We denote \textbf{ESS / ISM} (artificial retransmission) the test-set, where  clean signal  $s[t]$ was convolved with RIRs $h_{r,c}[t]$  either estimated by ESS or generated by the ISM method. The resulting speech signal is then given by standard convolution:  
\begin{equation}
  s_{r,c}[t] = s[t] * h_{r,c}[t] + \alpha n_{r,c}[ t+ \mathit{offset}], 
\end{equation}
In case noise $n_{r,c}[t]$ was added, the weight $\alpha$ is set to match the Signal-to-Noise Ratio $SNR(s_{r,c})$  estimated from the real retransmission condition in room $r$ and microphone $c$ using reference speech/non-speech segmentation and A-weighting function. The starting position $\mathit{offset}$ in the noise was selected randomly, then we repeated the noise in a loop to fill the whole audio (our noise samples are $1$ minute long). Data with added noise are  marked with  \textbf{noise} label. 

In generating RIRs using ISM, we did our best to be as close as possible to the real room setup (room dimensions, loudspeaker and microphone position, microphone orientation, $RT_{30}$ value). We estimated the $RT_{30}$ from logarithmic decay curve~\cite{Kuttruff2009} which was computed from an impulse response based on Schroeder integration~\cite{Schroeder1965}. $RT_{30}$ was applied in ISM method using Sabin-Franklin's formula~\cite{Pierce89}.

\subsection{Simulated (ISM) vs. real (ESS) RIRs}

\begin{figure}[!t]
\centering
\includegraphics[width=8cm]{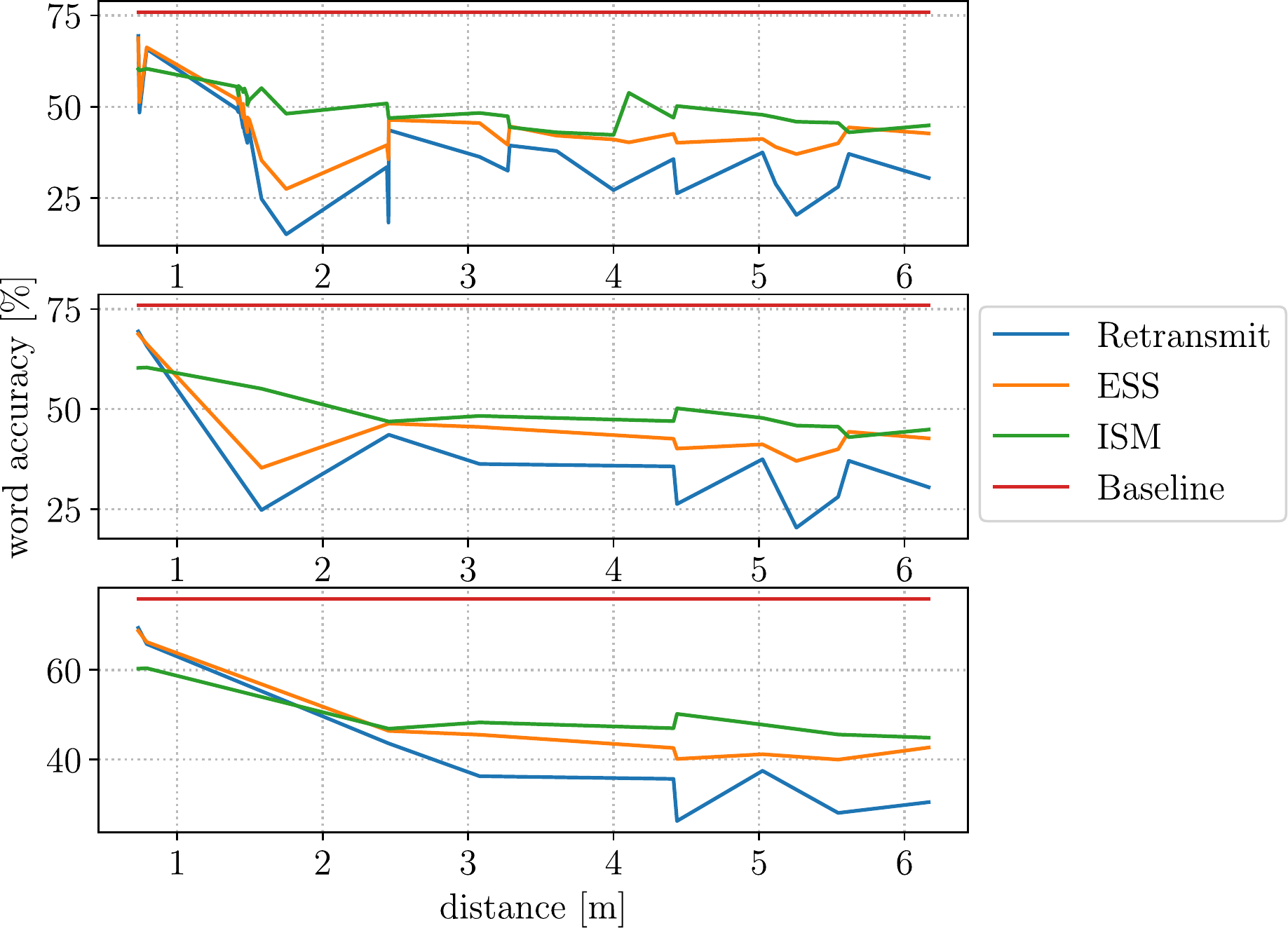}
\caption{Comparison of Real-Retransmitted, ESS Artificial-Retransmitted and ISM Artificial-Retransmitted test-sets in room L$207$. We sorted the microphones according to the distance from the loudspeaker (x-axis). The top panel shows all microphones. The middle panel shows only microphones in front of the loudspeaker ($\pm 90 \degree$). The bottom panel shows only microphones in front of the loudspeaker ($\pm 90 \degree$) with direct visibility.}
\label{fig_essism_distancel207}
\end{figure}

\begin{figure}[!t]
\centering
\includegraphics[width=8cm]{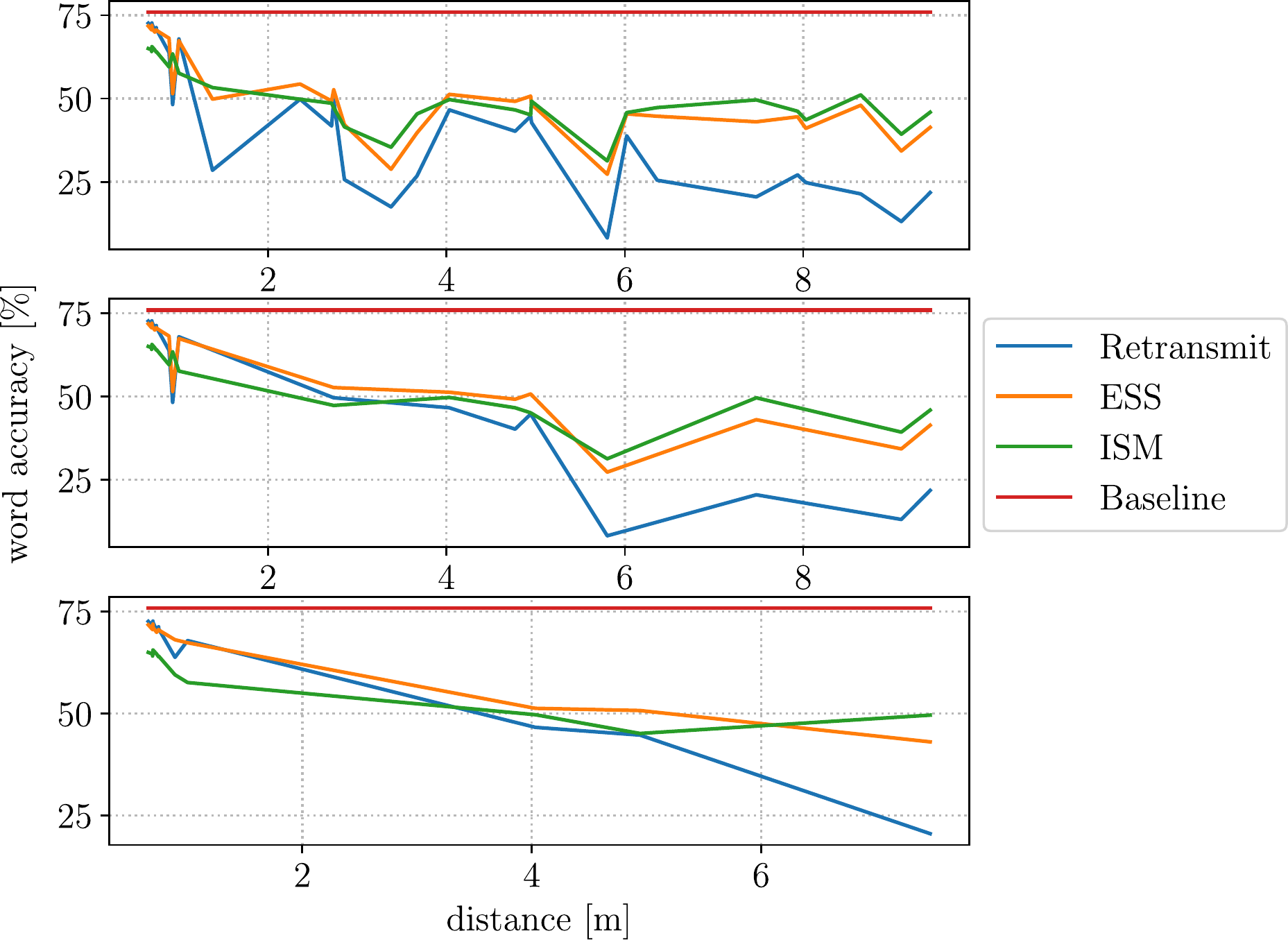}
\caption{Comparison of Real-Retransmitted, ESS Artificial-Retransmitted and ISM Artificial-Retransmitted test-sets in room Q$301$. We sorted the microphones according to the distance from the loudspeaker (x-axis). The top panel shows all microphones. The middle panel shows only microphones in front of the loudspeaker ($\pm 90 \degree$). The bottom panel shows only microphones in front of the loudspeaker ($\pm 90 \degree$) with direct visibility.}
\label{fig_essism_distanceq301}
\end{figure}

This section compares the influence of simulated RIR (ISM calculation) and real RIR (ESS estimation) on word accuracy. We used RIRs from two rooms and compared the Artificial-Retransmitted data to the Real-Retransmitted. As we can see from Figures~\ref{fig_essism_distancel207} and~\ref{fig_essism_distanceq301}, there is a gap between the Real-Retransmitted and both Artificial-Retransmitted data. This is caused by missing noise (see the following section). The ESS method provides slightly more realistic RIRs to the ISM, as the word accuracies are closer to the Real-Retransmitted data.

\subsection{Influence of noise on room acoustic simulation}
\begin{figure}[!t]
\centering
\includegraphics[width=8cm]{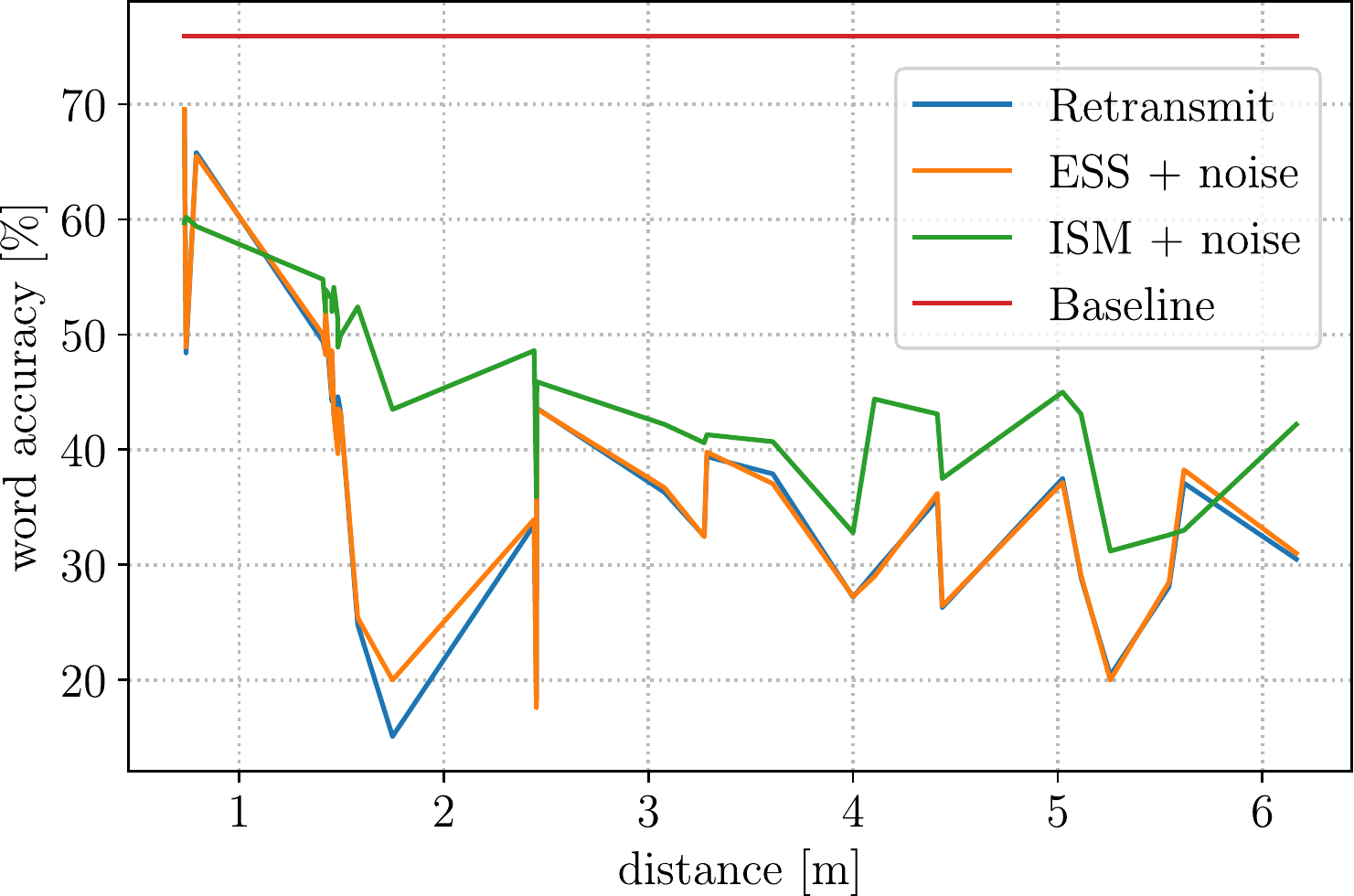}
\caption{Comparison of Real-Retransmitted, ESS Artificial-Retransmitted and ISM Artificial-Retransmitted test-sets in room L$207$. We sorted the microphones according to the distance form the loudspeaker (x-axis).}
\label{fig_essism_distance-noisyl207}
\end{figure}

\begin{figure}[!t]
\centering
\includegraphics[width=8cm]{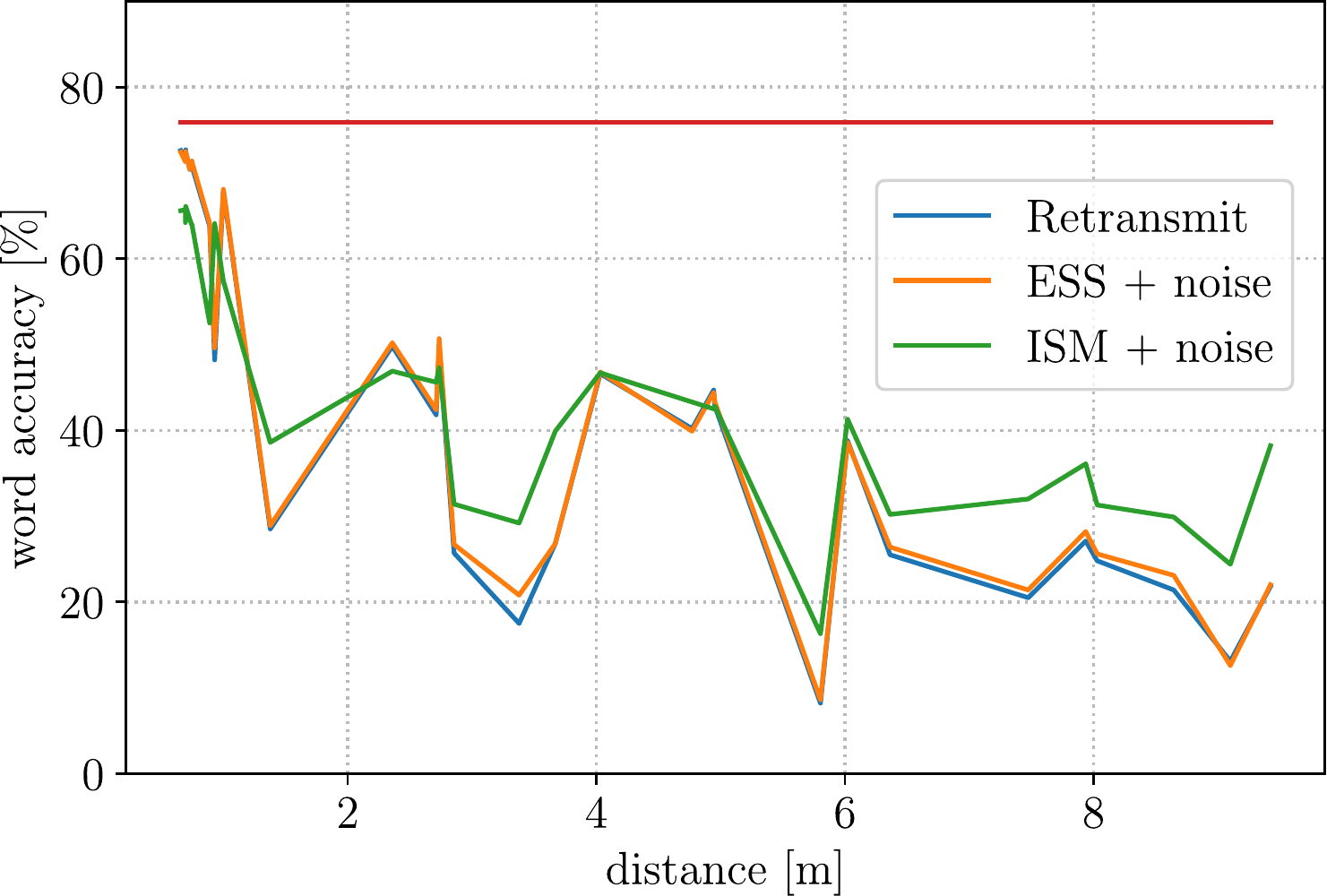}
\caption{Comparison of Real-Retransmitted, ESS Artificial-Retransmitted and ISM Artificial-Retransmitted test-sets in room Q$301$. We sorted the microphones according to the distance form the loudspeaker (x-axis).}
\label{fig_essism_distance-noisyq301}
\end{figure}

We show the need of noise for test data processing in this section. We use the same data setup as in the previous section and add noise. It is a matching noise, as it comes from the particular room and microphone as mentioned earlier. As we can see from Figures~\ref{fig_essism_distance-noisyl207} and~\ref{fig_essism_distance-noisyq301} compared to Figures~\ref{fig_essism_distancel207} and~\ref{fig_essism_distanceq301}, the gap between the Real-Retransmitted and ESS Artificial-Retransmitted data almost disappears. On the other hand, there is still a gap between ISM and ESS methods showing that the artificial RIR estimation is not good enough, especially for microphones placed in non-common positions (drawer, waste bin, book shelf, etc.).

\subsection{Microphones occlusion}\label{sec_mic_occlusion}
We analyzed how microphone occlusion impacts the WAC and the influence of the RIR estimation method. The impact is measured in WAC difference between Real-Retransmitted test-set and either an ESS Artificial-Retransmitted or ISM Artificial-Retransmitted test-set. We included both, only reverberation and reverberation plus additive noise. The results are shown in Table~\ref{tab_occlusion}. It is obvious that there is no significant difference between Real-Retransmitted and the ESS Artificial-Retransmitted test-set (measured on WAC) in all microphone placement conditions. Note however a stronger degradation of ISM compared to ESS Artificial-Retransmitted in occluded microphones.

\begin{table}[htb]
\setlength{\tabcolsep}{4.4pt}
\centering
\begin{tabular}{|l|c|c|c|c|}
\hline
\multicolumn{1}{|c|}{Microphone}               & \multicolumn{2}{|c|}{reverb}  & \multicolumn{2}{|c|}{reverb+noise}  \\ \cline{2-5}
\multicolumn{1}{|c|}{position}                       &    $RR-ISM$        &   $RR-ESS$          &   $RR-ISM$        &  $RR-ESS$       \\ \hline  
Face-to-face                                       & $  \ \,-1.6 \pm9.0 $ &  $  \ \,-4.0 \pm5.1 $ & $  \ \ \ \,1.3 \pm6.4 $ & $ -0.2 \pm0.5 $  \\ \hline
Partly boxed                                       & $ -19.8 \pm6.8 $ &  $ -13.0 \pm5.3 $ & $-13.0 \pm6.7 $ & $ -0.2 \pm0.5 $  \\ \hline
Fully boxed                                        & $ -21.5 \pm7.8 $ &  $ -11.9 \pm5.8 $ & $-14.8 \pm8.0 $ & $ -2.2 \pm2.2 $  \\ \hline
\end{tabular}
\vspace*{3mm}
\caption{Comparison of WAC differences between Real-Retransmited and ISM Artificial-Retransmitted test-sets (RR minus ISM) and Real-Retransmited and ESS Artificial-Retransmitted test-sets (RR minus ESS) using just reverberation (reverb) or reverberation and additive noise (reverb+noise). The differences are expressed as mean $\pm$ standard deviation --- a negative number means that ISM or ESS provides better results than those obtained with Real-Retransmited data. The data comes from rooms L$207$ and Q$301$ and the microphones have face-to-face orientation and direct visibility, are partly boxed (hidden in a shelf) and fully boxed (hidden in a drawer). It is obvious that for hidden microphones, only ESS RIR estimation with added noise provides meaningful test data.}
\vspace*{-5mm}
\label{tab_occlusion}
\end{table}

\section{ASR training data augmentation}\label{sec_amiexp}
In a real-world ASR,  one has to train ASR which is able to cope with a particular channel (far-field  microphone in our case) without having target  training data. As mentioned in Section~\ref{sec_relatedwork}, the best performing technique is data augmentation. We used an AMI dataset~\cite{McCowan2005} for this experiment; our unseen channel was the \emph{Single Distant Microphone} -- SDM and the only data available was \emph{Individual Headset Microphone} -- IHM. Our goal is to test data augmentation of AMI data using BUT ReverbDB  and to investigate suitable  reverberation techniques.  We do not run extensive experimentation  with noises; we use just the noises from BUT ReverbDB and add them to the training audio.

This set of experiments is inspired by Ko et al.~\cite{Ko2017}. Their work was aimed at comparing of real and simulated RIRs and adding point source noises to ASpIRE~\cite{Harper2015} and AMI datasets. On AMI, however, they only reported the impact of adding reverberated close-talk data (IHM) to the genuine distant microphone training data (SDM/MDM). We are not using SDM/MDM at all in the training.

We selected four BUT ReverbDB rooms  closest to AMI meeting rooms in type and dimensions as a source of real RIRs: Q301, L207, L212, and R112 (see Table~\ref{tab_rooms_measuredsofar}). We did not use other public RIR sources. We generated artificial RIRs similar to the four real rooms to compare artificial versus real RIRs. Theoretically, we can generate a large number of artificial RIRs with a good chance to hit the same room configuration (dimensions, reflection coefficients, speech source and microphone position) as the target data (AMI dataset). We consider this as cheating for the time being, but we would like to perform such an experiment in our future work. 

Each experiment is tagged with a used RIR set: artificial RIR (\textbf{AR}) or real RIR (\textbf{RR}) is accompanied with a number of RIRs used (\textbf{2k}, \textbf{306}, \textbf{30}). We add tag \textbf{ctXm} noting the microphone is in the range of $1$ to $X$ meters from the loudspeaker. \textbf{vis} denotes direct visibility between the microphone and the loudspeaker. Finally, \textbf{f2f} denotes ``face-to-face'' orientation of microphone and loudspeaker. In this way, \emph{*30.vis.ct2m.f2f} defines a set of $30$ RIRs, where microphones are directly visible, closer than $2$ meters and face-to-face oriented to the loudspeaker, and \emph{*306.vis.ct3m} defines a set of $306$ RIRs, where microphones are directly visible and closer than $3$~meters to the loudspeaker.

The training data augmentation was done in two steps: 1)~reverberating the IHM audio files using selected RIRs, and 2)~adding stationary noises to achieve SNR in the range of $10$~to $20$dB with uniform distribution. The reverberation was done in two ways:  either we convolved one whole audio file with one RIR, or changed the RIRs on-the-fly during convolution (see Section~\ref{sec_persegment_reverb} for details).

\subsection{Baseline system description}\label{sec_kaldirecip}
\begin{figure*}[!t]
\centering
\includegraphics[width=0.99\textwidth]{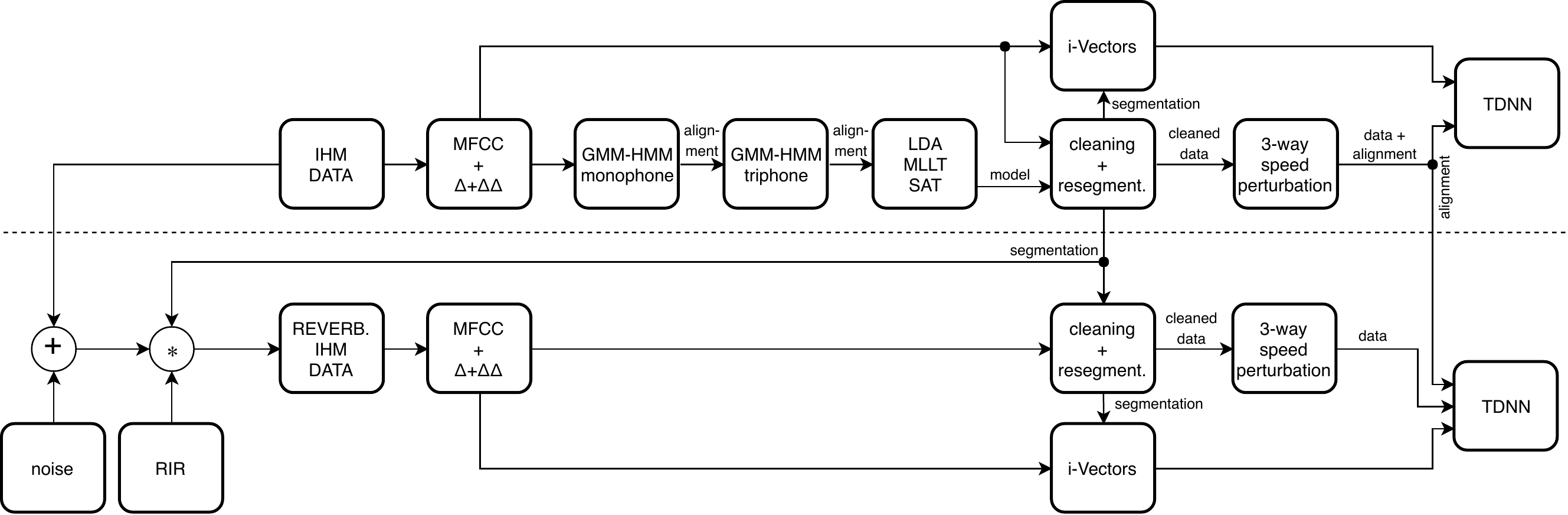}
\caption{Schemes of the baseline system (above the dashed line) and the modified system for reverberated data (below the dashed line).}
\label{fig_system_scheme}
\end{figure*}

For acoustic models training, we used a standard AMI recipe in Kaldi~\cite{Povey2011}. The baseline system is depicted in Figure~\ref{fig_system_scheme} above the dashed line. First, $13$-dimensional MFCC, delta and double-delta features are extracted.  Cepstral mean and variance normalization (CMVN) is performed. Mono-phone GMM-HMM model is trained on a subset of the training data (about $10.8$ hours of AMI IHM audio). All the data is then aligned using this system. Context-dependent tri-phone model training on the full training set  (about $78$ hours of audio) follows, and the data is re-aligned. Further, features are spliced together, projected to $40$-dimensional space using linear discriminant analysis (LDA), and a de-correlation based on the maximum likelihood linear transform (MLLT) is applied. In the last step, the model is retrained using speaker adaptive training (SAT). The training data is re-segmented and only the audio matching the transcriptions is selected (cleaning process) based on decoding with the GMM-HMM model and   biased language model  built from a reference transcript. In this way, about $7$~hours of audio are discarded from the full training set. After this, the cleaned full training set is speed perturbed (original plus two speed alternations) resulting in about $210$ hours of training audio. The state alignments generated by GMM-HMM system are used for DNN training. The DNNs are trained on $40$-dimensional filter-bank energies along with $100$-dimensional i-Vectors~\cite{PeddintiIVecs2016}. A time delayed neural network (TDNN) trained with lattice-free MMI objective  is used as the  final acoustic model. 

\subsection{Modifications of Kaldi baseline}\label{sec_sysmod}
The standard AMI recipe uses the training data both for cleaning and segmentation, and for the actual acoustic model training. When using reverberated data for all these steps, we found a significant decrease in accuracy (caused obviously by worse  models) and fluctuations in the amount of retained audio. Therefore, we decided to ``freeze'' the baseline  system segmentation across all further experiments, which also implies that the same amount of training data was used (210 hours with speech perturbation). The segmentation also served for i-Vector resets (see below in Section~\ref{sec_persegment_reverb}). In the same manner, we also consistently used the baseline system alignment to train all DNN acoustic models. The modifications of a baseline system for the reverberated data are depicted in Figure~\ref{fig_system_scheme} below the dashed line.

\subsection{Averaging results}
\begin{figure}[!t]
\centering
\includegraphics[width=0.45\textwidth]{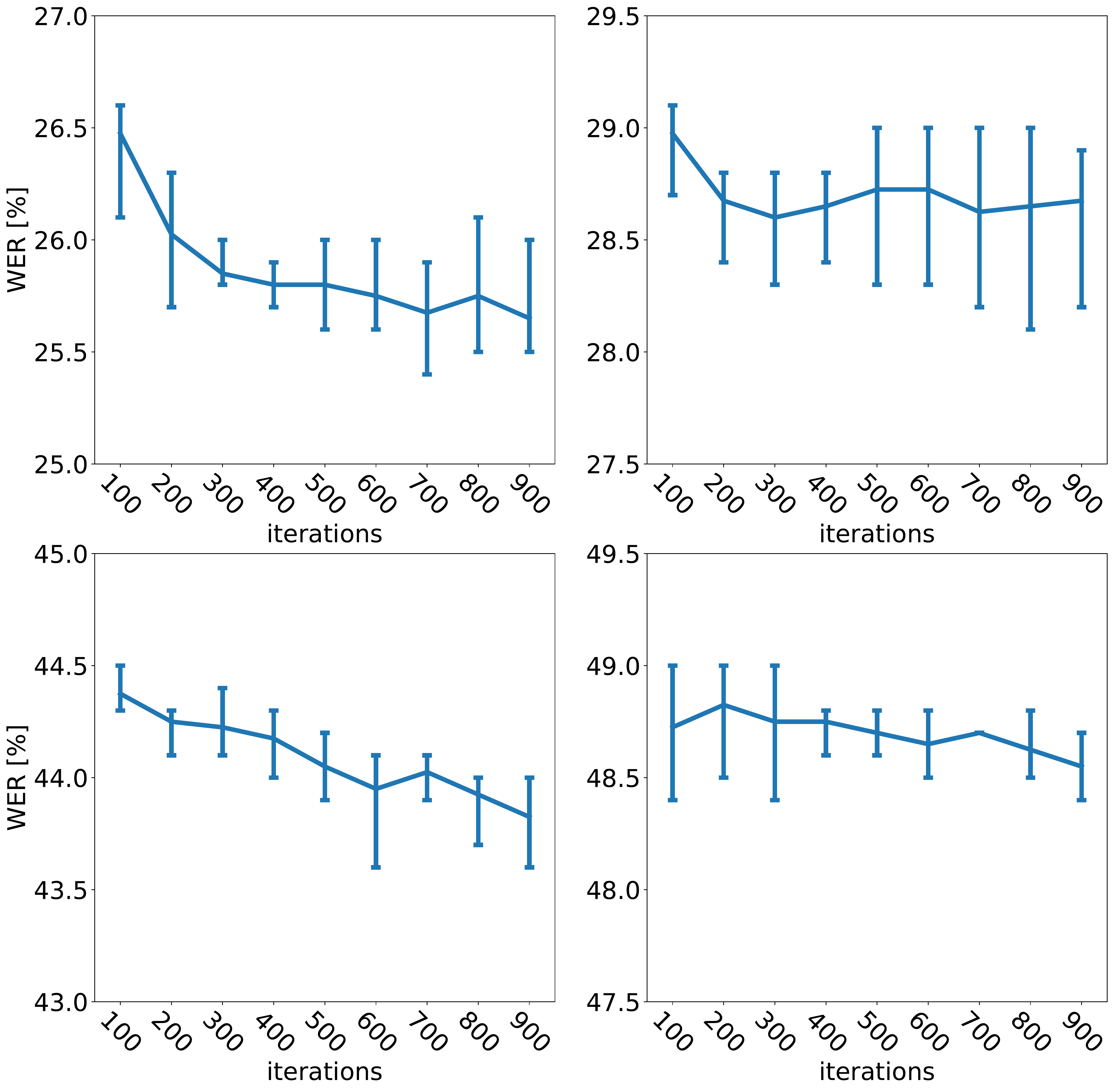}
\caption{Comparison of mean WER (over $5$ runs). X-axis is the number of iterations in training NN, Y-axis is achieved WER for IHM (top row), SDM (bottom row), dev (left column), and eval (right column) sets. The solid left-to-right line connects means, the top and bottom lines show maximum and minimum WERs achieved for a particular run.}
\label{fig_tdnns_iters}
\end{figure}

When we experimented with Kaldi AMI recipe, we found that the resulting WER in not stable enough\footnote{This is a known issue of Kaldi, probably caused by inherent nondeterminism of GPU-based matrix multiplication, as discussed by Kaldi core developers at https://github.com/kaldi-asr/kaldi/issues/2905.}. When  an experiment was  run several times, we observed WER fluctuations in tenths of percent. Stability does not improve when adding more NN training iterations. As some of our experiments also differ in tenths of percents, our conclusions would not be  statistically  significant. That is why all results presented in this section are averages over $5$ runs of ASR training (see Figure~\ref{fig_tdnns_iters} for details). We performed Student's T-test on selected pairs of systems with close average results. We concluded that $0.2\%$ absolute difference on WER for the significance level $\alpha = 0.05$ is statistically significant ($0.1\%$ absolute difference is not significant).

\subsection{Per segment reverberation}\label{sec_persegment_reverb}
The problems of the AMI dataset are long recording and relatively small number of speakers ($547$). So even if we generate thousands of RIRs using ISM, only $547$ are used if we apply one RIR on one whole audio file.
The AMI recipe contains speaker adaptation using i-Vectors~\cite{PeddintiIVecs2016}. Each i-Vector is estimated on-the-fly on $2-10$ speech segments and then it is reset to ensure data variability and to prevent TDNN over-training. We modified our reverberation algorithm in order to allow changes of RIR during convolution with the audio. In the end, every speaker is reverberated with a set of RIRs and the data variety is increased compared to a single audio file reverberation.

The results (Table~\ref{tab_perspk_reverb}) show that bringing more environmental variability per i-Vector, the reverberation decreases WER from $43.42\% / 48.46\%$ to $41.80\% / 47.06\%$ for SDM dev / eval set. We also  conducted an experiment, where we changed the RIR only in silences longer than $3$ seconds, in order to prevent artifacts in the convolution, as the RIR is $1$ second long. This also ``stabilizes'' the channel for the i-Vector extraction and makes the i-Vectors focus on the speaker rather than acoustic environment. Here we obtained another slight WER decrease from $41.80\% / 47.06\%$ to $41.70\% / 46.74\%$ on SDM dev / eval.

\begin{table}[htb]
\centering
\begin{tabular}{|l|c|c|c|c|c|}
\hline
\multicolumn{1}{|c|}{\multirow{3}{*}{System}}      & \multirow{3}{*}{Segm \#} & \multicolumn{4}{|c|}{WER [\%]}  \\ \cline{3-6}
                                                   &         & \multicolumn{2}{|c|}{IHM} & \multicolumn{2}{|c|}{SDM} \\ \cline{3-6}
                                                   &         & dev    & eval   &  dev    & eval    \\ \hline  
ihm.AR2k.ct3m.insil                                & 36357   & 	21.52 &	23.06  & 41.70   &	46.74  \\ \hline
ihm.AR2k.ct3m.per1seg                              & 33312   & 	21.44 &	23.02  & 41.80   &	47.06 	\\ \hline
ihm.AR547.ct3m.perfile                              & 547     &  21.72 &	23.24  & 43.42   &  48.46  \\ \hline
\end{tabular}
\vspace*{3mm}
\caption{Comparison of ``per segment''  with  ``per file'' reverberation.  \emph{per1seg} setup changes RIR  in synchrony  with Kaldi i-Vector speaker adaptation. \emph{insil} denotes experiment, where RIR is changed only in silences longer than $3$ seconds.  Column \emph{Segm \#}  shows numbers of segments with fixed RIR.
We randomly draw $547$ RIRs from $2000$ set for ihm.AR547.ct3m.perfile system.}
\vspace*{-5mm}
\label{tab_perspk_reverb}
\end{table}

\subsection{Room impulse response passivation and delay compensation}\label{sec_passivate_shift}

Having estimated real or generated artificial RIRs, one may post-process them to achieve more consistent results and to overcome over-excitation and delays caused by the convolution.
The delay in any RIR is caused by the speed of sound  and can be partly compensated by measurement of microphone to loudspeaker distance. However, precise  compensation is hard due to humidity and air pressure changes. Delay compensation in ISM synthesis of RIR is theoretically straightforward; the delay can be computed analytically. On the other hand, it may produce an incorrectly delayed RIR in the case of a cardioid microphone and the sound source are placed exactly behind the microphone. In  this case, the direct signal is zero and we see only the reflections with larger delay than we expect from the microphone--loudspeaker distance and the speed of sound. 
The delay compensation is critical in data augmentation for ASR training~\cite{karafiat2017,Ko2017}. First, the labels (phonemes, senones, etc.) are aligned with the training ``clean'' speech data using a decoder. The clean data are swapped with the augmented (reverberated) version in the next step of training. Here, the original alignment (timing) is used with the augmented data and any time shift caused by RIR delay leads to label versus data mismatch. We denote systems with applied delay compensation by tag~\textbf{shi}.

\begin{table}[htb]
\centering
\begin{tabular}{|l|c|c|c|c|}
\hline
\multicolumn{1}{|c|}{\multirow{3}{*}{System}}      & \multicolumn{4}{|c|}{WER [\%]}  \\ \cline{2-5}
                                                   & \multicolumn{2}{|c|}{IHM} & \multicolumn{2}{|c|}{SDM} \\ \cline{2-5}
                                                   & dev    & eval   &  dev    & eval    \\ \hline \hline 

ihm.AR2k.vis.ct3m.perfile                          & 25.80  & 28.40  &  44.38  & 48.48   \\ 
ihm.AR2k.\textbf{pas}.vis.ct3m.perfile             & 21.72  & 23.24  &  43.42  & 48.46   \\ \hline

ihm.RR306.vis.ct3m.perfile                          & 25.83  & 28.35  &  44.05  & 48.55   \\ 
ihm.RR306.\textbf{pas}.vis.ct3m.perfile             & 25.44  & 28.42  &  44.14  & 48.54   \\ \hline \hline 

ihm.AR306.pas.vis.ct3m.per1seg                      & 21.40  & 23.00  &  41.72  & 46.76   \\ 
ihm.AR306.pas.\textbf{shi}.vis.ct3m.per1seg         & 21.46  & 23.35  &  41.78  & 47.30   \\ \hline
        
ihm.RR306.pas.vis.ct3m.per1seg                      & 25.22  & 27.18  &  43.26  & 47.42   \\ 
ihm.RR306.pas.\textbf{shi}.vis.ct3m.per1seg         & 25.32  & 27.72  &  43.10  & 47.12   \\ \hline

ihm.RR30.pas.vis.ct2m.f2f.per1seg                   & 23.12  & 24.80  &  42.30  & 46.36   \\ 
ihm.RR30.pas.\textbf{shi}.vis.ct2m.f2f.per1seg      & 22.88  & 24.42  &  42.42  & 46.44   \\ \hline

\end{tabular}
\vspace*{3mm}
\caption{Comparison of the effect of passivation \textbf{pas} -- top panel, and delay shift \textbf{shi} -- bottom panel, on various RIR sets. }
\vspace*{-5mm}
\label{tab_passiveshift_reverb}
\end{table}

Another problem is over-excitation caused by amplifying the audio using a RIR, leading to signal clipping. To overcome this, we scale the RIR to a level which ensures that no single magnitude in the frequency response is larger than $1$.
The passivation has no effect when using floating-point arithmetic for convolution followed by a normalization. On the other hand, one may still face fixed-point implementations/scenarios where (latent) an overflow has a significant impact. We denote systems with applied passivation by tag \textbf{pas}. 

We summarized results with RIR passivation and delay shift in Table~\ref{tab_passiveshift_reverb}. Passivation experiments are shown in the first four lines, both for artificial and real RIRs. We can clearly conclude that passivation significantly helps for artificial RIRs in the IHM condition. Passivation does not bring any significant improvement for real RIRs. This leads to the conclusion that ReverbDB RIRs are well estimated and will not cause over-excitation compared to ISM generated RIRs which may cause signal clipping. Our finding is that ISM-generated RIRs often lead to over-excitation. As the IHM data contains strong audio signals, in combination with the amplifying ISM RIR, the augmented training data is heavily clipped. This leads to overall ASR system degradation. ESS RIRs do not have this issue. 

The last four lines aim at RIR delay compensation. When analyzing the distribution of delays, we found, that artificial RIRs have a peak at $0$ seconds with about $1/4$ of them uniformly distributed from $0$ to $0.02$ seconds ($2$ frames). On the other hand,  real RIRs delay distribution is Gaussian with peak at $0$ and tailing to $\pm 0.05$ seconds with extreme values reaching  $0.14$ second ($14$ frames). A negative delay can be caused, for example, by less precise loudspeaker to microphone distance measurement. A small positive delay is not so substantial as it only leads to delaying the reverberated audio with respect to the alignment, and a delay within $1-2$ frames can be considered as wanted variability. Larger delays may cause degradation due to desynchronization of the audio and alignment in NN training (see Section~\ref{sec_sysmod}). However, a negative delay is critical, as when we try to compensate it, the beginning of RIR (containing the important direct sound and early reflections!) is cut off. Such trimmed RIR is damaged, as it does not carry full  information on the acoustic environment anymore. 

The results (lines $5$ and $6$ in Table~\ref{tab_passiveshift_reverb}) show that applying delay compensation (synchronizing all RIRs to start at $0$~seconds) for artificial RIRs does not have significant impact except for small deterioration for SDM eval set. Applying delay compensation for real RIRs (lines $7$ to $10$ in Table~\ref{tab_passiveshift_reverb}) has mixed results. Small errors in distance measurement can actually  bring wanted variability to the augmented data in some cases. We decided to use passivation but not delay compensation in further experiments, as the former has clearly gain, but the results of the later can be considered as statistical noise.

\subsection{Simulated versus real room impulse responses on AMI data}

\begin{table}[htb]
\centering
\begin{tabular}{|l|c|c|c|c|}
\hline
\multicolumn{1}{|c|}{\multirow{3}{*}{System}}      & \multicolumn{4}{|c|}{WER [\%]}  \\ \cline{2-5}
                                                   & \multicolumn{2}{|c|}{IHM} & \multicolumn{2}{|c|}{SDM} \\ \cline{2-5}
                                                   & dev    & eval   &  dev    & eval    \\ \hline \hline
ihm (\textbf{baseline})                                     & 20.02  & 20.04  &  60.12  & 72.70   \\ \hline \hline

ihm.AR2k.pas.vis.ct3m.per1seg              & 21.44  & 23.02  &  41.80  & 47.06   \\ \hline

ihm.RR306.pas.vis.ct3m.per1seg             & 25.22  & 27.18  &  43.26  & 47.42   \\ \hline

ihm.AR306.pas.vis.ct3m.per1seg             & \textbf{21.40}  & \textbf{23.00}  &  \textbf{41.72}  & 46.76   \\ \hline

ihm.RR30.pas.vis.ct2m.f2f.per1seg          & 23.12  & 24.80  &  42.30  & \textbf{46.36}   \\ \hline

ihm.AR30.pas.vis.ct2m.f2f.per1seg          & 21.86  & 23.70  &  41.92  & 46.76   \\ \hline\hline

ihm.RR306.pas.vis.ct3m.per1seg +           &        &        &         &         \\
ihm.AR306.pas.vis.ct3m.per1seg             & 22.30 	& 23.90  & 41.80   &  46.24     \\ \hline

ihm.RR30.pas.vis.ct2m.f2f.per1seg +        &        &        &         &         \\
ihm.AR2k.pas.vis.ct3m.per1seg              & 21.86   & 23.22   &  \textbf{41.54}   & \textbf{46.12}     \\ \hline \hline

sdm1 (\textbf{target})                     & 29.38   & 36.74   &  35.72   & 39.65    \\ \hline
\end{tabular}
\vspace*{3mm}
\caption{Comparison of various ASRs trained on augmented IHM to the baseline (ASR train on clean IHM) and ``cheating'' target (ASR trained on SDM) systems. The bottom part compares system combination (on training data level).
}
\vspace*{-5mm}
\label{tab_realvsartif_reverb}
\end{table}

We compared the influence of artificial RIRs (ISM generated) with real RIRs (estimated  from BUT ReverbDB using ESS method) in the following experiments. It should be remembered, that the scenario is training ASR to target an unseen environment (AMI meeting rooms) without having any target data. We tried to answer the following questions:
\begin{itemize}
\item How many RIRs are sufficient?
\item Are artificial RIRs superior to real ones?
\item Are artificial and real RIRs complementary?
\end{itemize}

We summarized our results in Table~\ref{tab_realvsartif_reverb}. The baseline system \textbf{ihm} is trained on AMI IHM data using default Kaldi recipe (Section~\ref{sec_kaldirecip}). This system performs well on in-domain IHM dev ($20.02\%$ WER) and eval ($20.04\%$ WER) data, but very badly on target (out-of-domain) SDM dev ($60.12\%$ WER) and eval ($72.70\%$ WER). To have an idea of the best reachable WER, we trained the system on target data -- \textbf{sdm1}, in the same way  as the \textbf{ihm}. We achieved expected huge improvement on (in this case in-domain) SDM dev ($35.7\%$) and eval ($39.6\%$) data, but significant deterioration on (now out-of-domain) IHM dev ($29.3\%$) and eval ($36.7\%$) data.

We then applied various data augmentation techniques on IHM training data to simulate the target environment and to achieve an ASR adapted to SDM data, without seeing any SDM data. We use the following notation:
\begin{itemize}
\item {\bf RR30} -- set of $30$ real RIRs including $30$ microphones from $4$ rooms of BUT Reverb DB (see Section~\ref{sec_reverbdb}) with microphones in a range of $1$-$2$ meters from the loudspeaker and face-to-face orientation.
\item {\bf RR306} -- set of $306$ real RIRs including $306$ microphones from the $4$ rooms with microphones in a range of $1 - 3$ meters from the loudspeaker and direct visibility. \emph{RR306} is superset of \emph{RR30}.
\item {\bf AR30} -- set of $30$ artificially generated RIRs with microphones in a range of $1 - 2$ meters from the loudspeaker and face-to-face orientation. This set is a random draw from a larger set of artificial RIRs with parameters set to as close as possible to the $4$ rooms. This set should be comparable to \emph{RR30}.
\item {\bf AR306} -- set of $306$ artificially generated RIRs with microphones in a range of $1 - 3$ meters from the loudspeaker and direct visibility. This set is a random draw from \emph{AR2k} set
\item {\bf AR2k} -- set of $2000$ artificially generated RIRs with microphones in a range of $1 - 3$ meters from the loudspeaker and direct visibility. Parameters of the RIRs are as close as possible to the $4$ rooms.
\end{itemize}

By comparing the results of five  systems from the upper part of Table~\ref{tab_realvsartif_reverb}, we can conclude that using a larger set of RIRs is not always beneficial --- see the significant gain when going from \emph{AR30} to \emph{AR306}, but no gain or even deterioration when going from \emph{AR306} to \emph{AR2k} and a significant deterioration for real RIRs -- going from \emph{RR30} to \emph{RR306}. We conclude that a careful selection of RIRs covering the target scenario is important.

Comparing the artificial RIR (\emph{AR*}) to real RIR (\emph{RR*}) systems shows no clear winner. Artificial RIRs have a significant advantage in working well on IHM data too, making the ASR more robust on both IHM and SDM data. On the other hand, \emph{ihm.RR30.pas.vis.ct2m.f2f.per1seg} system is significantly better on the SDM eval set.

Finally, artificial and real RIRs seem to be complementary and their combination is beneficial (see bottom part of Table~\ref{tab_realvsartif_reverb}). The  combination was done  on the level of training data by taking one half of data augmented by artificial RIRs and one half of data augmented by real RIRs (in order to always train on the same amount of data). \emph{RR30 + AR2k} achieved the best WER on the SDM data set with small deterioration on IHM data set compared to the best single systems.

\section{Conclusions and future work}\label{sec_conclusion}
This paper presents BUT ReverbDB, a public set  of  RIRs, noise and retransmitted data for ASR and SRE development and testing. The set is available for free under a non-restrictive CC-BY license, and covers non-standard positions of microphones that are interesting  for investigation/intelligence scenarios.Currently, the set contains data from $8$ rooms and will continue to grow. We believe that our paper can serve as a cook-book of how to collect such dataset.

A set of experiments aiming at the ASR test data processing was performed in order to check and validate the database, with interesting findings: Clock asynchronicity problem in RIR estimation by MLS technique was studied and we found that it can be fixed by estimating the clock ratio using cross-correlation (when applied, we obtained  comparable WER results  as with the ESS technique). We also confirmed other papers' conclusion on the importance of  adding real noise in ASR test data preparation. Finally, we observed a clear superiority of real RIRs over artificial ones.

ASR training data augmentation experiments targeted training of an ASR system on data augmented by real or artificial RIRs. We have found the passivation of RIR is extremely important, and recommend checking this issue in other RIR datasets. We also concluded that knowing the target room configuration is beneficial, as we obtained  better results with a few carefully selected RIRs than with a huge number of randomly picked ones. In real applications, this  calls for a system capable of  extracting  RIRs from reverberated audio and its use for the augmentation of training data.  
We have also shown that real and artificial RIRs are  complementary, and investigated into a number of technical (but nonetheless important) issues such as reverberation of long audio files per  speaker, and RIR delay compensation. 

In future work, we would like to grow our data-set, and extend it  with real speech data. Our experimental work will include  investigation into the influence of having just one or two  IRs from one room rather than many IRs from one room, a simplification of ASR system (i.e. producing results without  i-Vector adaptation) and also changing the noise within each speaker adaptation segment.

\appendices

\section{Measuring RIRs in BUT ReverbDB}\label{sec_ReverbDB_appendix}

This section contains a more detailed description of the hardware and software used and metadata collected. Even more details accompanied with photos are available in the technical report which is part of the BUT ReverbDB release\footnote{\url{https://speech.fit.vutbr.cz/software/but-speech-fit-reverb-database}}.
\label{sec_measuringReverbDB}
\subsection{Hardware}\label{sec_HW}

\subsubsection{Audio recording}

Our requirements on recorded audio are a large amount of channels in high quality and sample-to-sample synchronization across all channels (see Section~\ref{sec_howtogetRIR}) at reasonable price\footnote{We are aware of AVS or other hi-end solutions with master/slave clock bus etc., but these  were unacceptable for our budget.}. We decided to design our own hardware with the help of colleagues from Audified\footnote{\url{http://www.audified.com}}. The device is based on Analog Devices development board SC589 equipped with an ARM Cortex A5 processor and Sharc DSP processor. The processor board is connected to two $16$-channel boards equipped with $96$kHz, $24$bit, AKM A/D converters with software driven gains and phantom power. 
The sampled audio data are assembled  in TCP/IP packets (interleaving format with timestamps) and sent through Ethernet to a local recording device. The $32$ channels here are reconstructed and stored on a hard-drive as $32$ PCM audio files. Any packet drop-outs are reported into a log file.

\subsubsection{Audio playback}
We used an external USB stereo sound-card with symmetrical outputs. We played our audio data in the left channel together with a control signal played in the right channel.
The control signal allows us to detect possible problems (caused by a playback buffer under-run, samples drop, packet loss, etc.) and to split the long raw recordings back to the retransmitted audio corpus (parallel corpus).

The control signal is recorded as  channel $32$ on the recording device. The left channel is fed  to the loudspeaker -- Adam audio A$7$X studio monitor\footnote{RMS: $100$W, Frequency response: $42$Hz -- $50$kHz, Crossover frequency: $2.5$Hz, Size: $337$mm x $201$mm x $280$mm, Weight: $9.2$kg, Bass reflex in front. \url{http://www.adam-audio.com/en/pro-audio/products/a7x/description}}. 
The loudspeaker is placed on a wheeled stand with a settable height. The loudspeaker is placed in the following positions in each room:
\begin{itemize}
  \item Sitting person: Usually in front of a computer monitor or a table simulating a sitting person (about $140$cm above the floor).
  \item Standing person: Placed randomly in the room where a person can stay (about $170$cm above the floor).
  \item Noise source: Simulates position of a source of noise, for example a radio, air-condition (AC), fan, etc. The reasoning is to collect RIR of noise source and then generate ``real'' noises by,  for example,   reverberating an FM radio audio stream using this RIR.
  \item Non-standard position: Directed to the ceiling, or floor, lying on the floor, etc.
\end{itemize}

\subsubsection{Noise sources}
Most environments are without any additional noise source. The real noises include AC, vents, or common street noise coming through windows. We added artificial noise sources in a few recording sessions. This is marked in the  meta-data. We used a Tecsun PL-680 radio receiver tuned to a random local FM station as another source of noise. This noise source is placed in the usual radio positions in the room. 

\subsection{Microphones, mountings and positions}\label{sec_microphones}
We use two types of microphone capsules,  both with symmetrical wiring and phantom-powered: 
\begin{itemize} 
\item \emph{standard microphone capsule} (a majority of our microphones) includes an  omnidirectional electret condenser microphone module -- \texttt{PMOF-6027PN-42UQ}. 
\item \emph{Sennheiser} MKE 2 omnidirectional microphone. 
\end{itemize}
They are placed in several mountings:

\subsubsection{Spherical array mounting}
In order to cover the microphone array use-case, we designed a spherical $8$-channel array.
It consists of $8$ \emph{standard microphone capsules} placed in an $8$cm diameter sphere on two parallel planes ($4$ per each). Microphones are placed in square vertices. The two vertices are rotated by $45\degree$. The orientation of the microphones is from the sphere center. This microphone mounting is usually placed where a similar device (a smart home assistant) is expected in the room.

\subsubsection{Internet-of-Things mounting}
We mounted two standard microphone capsules into plastic boxes with magnets glued on. These devices are usually attached to a wall or some metal object mounted on a wall.
\subsubsection{Stand mounting}
$6$ to $10$ microphones are mounted on a stand.
These are then placed on floor, table, etc. and adjusted to  desired microphone position and direction. Some of the microphones are also mounted to a computer monitor, lamp, and other objects simulating  table-top microphones.

\subsubsection{Laid mounting}
$5$ to $10$ microphones are just laid on a chair, table, cupboard, shelf, etc. The microphone is usually oriented approximately towards the sound source.

\subsubsection{Hidden laid mounting}
Some of the laid microphones are partly or fully hidden in an object (occluded microphones). This simulates the placing of ``bugs'' and listening devices. The place is described in the particular microphone placement meta-data. We hid the microphone in a shelf, drawer, waste bin, flower, vent or behind painting, white board, etc.

\subsubsection{In-air mounting}
About $5$ microphones are placed in the air close to the ceiling. We use fishing rods here to place the microphones in the upper corners, close to various sensors (smoke detectors), lights, projectors, etc. We also let one or two microphones just hang down and be in the space far from any obstacles.

\subsection{Meta-data}
We generated a lot of meta-data to provide details on: $1$) the room (environment), $2$) loudspeaker placing(s), $3$) microphones placings.
The meta-data is available in the text files. We provide several coordinate systems allowing for easy work with our data set. We use absolute and relative Cartesian (depth, width, height) and spherical (distance, azimuth, elevation) coordinates for microphone and loudspeaker positions. We also use azimuth and elevation for microphone or loudspeaker orientation.

The origin for \emph{relative measurements} is the placement of the loudspeaker (speech source).
Microphone to loudspeaker distance can, therefore, be easily obtained by looking to the relative distance of the microphone.
In addition to the size of the room, we store photos, description, type, size, temperature, materials, amount of furniture, and background noise level.

We can place several microphone setups in every room, however, we use mainly just one microphone setup per room.
On the other hand, we usually place the loudspeaker(s) in several positions for every microphone setup. We try to have at least five distinct positions here. The first position of the loudspeaker is the one we use for measuring the coordinates of all microphones and their meta-data. One loudspeaker setup can consists of one or more physical loudspeakers. The first loudspeaker is always the one playing the audio (speech data, sine sweeps, MLS, etc.). The others may be used as noise sources (radio in the background etc.). We store coordinates (position), orientation (facing), and the type of the loudspeaker as meta-data.

\ifCLASSOPTIONcaptionsoff
  \newpage
\fi

\bibliographystyle{IEEEtran}
\bibliography{IEEEabrv,bibliography}

\begin{thebibliography}{10}
\providecommand{\url}[1]{#1}
\csname url@samestyle\endcsname
\providecommand{\newblock}{\relax}
\providecommand{\bibinfo}[2]{#2}
\providecommand{\BIBentrySTDinterwordspacing}{\spaceskip=0pt\relax}
\providecommand{\BIBentryALTinterwordstretchfactor}{4}
\providecommand{\BIBentryALTinterwordspacing}{\spaceskip=\fontdimen2\font plus
\BIBentryALTinterwordstretchfactor\fontdimen3\font minus
  \fontdimen4\font\relax}
\providecommand{\BIBforeignlanguage}[2]{{%
\expandafter\ifx\csname l@#1\endcsname\relax
\typeout{** WARNING: IEEEtran.bst: No hyphenation pattern has been}%
\typeout{** loaded for the language `#1'. Using the pattern for}%
\typeout{** the default language instead.}%
\else
\language=\csname l@#1\endcsname
\fi
#2}}
\providecommand{\BIBdecl}{\relax}
\BIBdecl

\bibitem{karafiat2017}
M.~Karafi{\'{a}}t, K.~Vesel{\'{y}}, K.~{\v{Z}}mol{\'{i}}kov{\'{a}},
  M.~Delcroix, S.~Watanabe, L.~Burget, J.~{\v{C}}ernock{\'{y}}, and
  I.~Sz{\H{o}}ke, \emph{\BIBforeignlanguage{english}{Training Data Augmentation
  and Data Selection}}, ser. Computer Science, Artificial Intelligence.\hskip
  1em plus 0.5em minus 0.4em\relax Springer International Publishing, 2017, pp.
  245--260.

\bibitem{Melot2015}
J.~Melot, N.~Malyska, J.~Ray, and W.~Shen, ``{Analysis of Factors Affecting
  System Performance in the ASpIRE Challenge},'' in \emph{2015 IEEE Workshop on
  Automatic Speech Recognition and Understanding (ASRU)}, Dec 2015, pp.
  512--517.

\bibitem{Mosner2018a}
L.~Mo{\v{s}}ner, O.~Plchot, P.~Mat{\v{e}}jka, O.~Novotn{\'{y}}, and
  J.~{\v{C}}ernock{\'{y}}, ``{D}ereverberation and {B}eamforming in {R}obust
  {F}ar-{F}ield {S}peaker {R}ecognition,'' in \emph{Proceedings of
  {I}nterspeech 2018}, vol. 2018, no.~9.\hskip 1em plus 0.5em minus 0.4em\relax
  International Speech Communication Association, 2018, pp. 1334--1338.

\bibitem{Mosner2018b}
L.~Mo{\v{s}}ner, P.~Mat{\v{e}}jka, O.~Novotn{\'{y}}, and
  J.~{\v{C}}ernock{\'{y}}, ``{D}ereverberation and {B}eamforming in
  {F}ar-{F}ield {S}peaker {R}ecognition,'' in \emph{Proceedings of {ICASSP}
  2018}.\hskip 1em plus 0.5em minus 0.4em\relax IEEE Signal Processing Society,
  2018, pp. 5254--5258.

\bibitem{Panayotov2015}
V.~Panayotov, G.~Chen, D.~Povey, and S.~Khudanpur, ``Libri{S}peech: {A}n {ASR}
  {C}orpus based on {P}ublic {D}omain {A}udio {B}ooks,'' in \emph{2015 IEEE
  International Conference on Acoustics, Speech and Signal Processing
  (ICASSP)}, April 2015, pp. 5206--5210.

\bibitem{Greenberg2010}
C.~Greenberg, A.~Martin, D.~Graff, L.~Brandschain, and K.~Walker, ``2010 {NIST}
  {S}peaker {R}ecognition {E}valuation {T}est {S}et {LDC}2017{S}06,'' Tech.
  Rep., 2017.

\bibitem{Yoshioka2012}
T.~Yoshioka, A.~Sehr, M.~Delcroix, K.~Kinoshita, R.~Maas, T.~Nakatani, and
  W.~Kellermann, ``{Making Machines Understand Us in Reverberant Rooms:
  Robustness Against Reverberation for Automatic Speech Recognition},''
  \emph{IEEE Signal Processing Magazine}, vol.~29, no.~6, pp. 114--126, Nov
  2012.

\bibitem{Harper2015}
M.~Harper, ``{The Automatic Speech recognition In Reverberant Environments
  (ASpIRE) challenge},'' in \emph{2015 IEEE Workshop on Automatic Speech
  Recognition and Understanding (ASRU)}, Dec 2015, pp. 547--554.

\bibitem{Janin2003}
A.~Janin, D.~Baron, J.~Edwards, D.~Ellis, D.~Gelbart, N.~Morgan, B.~Peskin,
  T.~Pfau, E.~Shriberg, A.~Stolcke, and C.~Wooters, ``{The ICSI Meeting
  Corpus},'' in \emph{2003 IEEE International Conference on Acoustics, Speech,
  and Signal Processing, 2003. Proceedings. (ICASSP '03).}, vol.~1, April 2003,
  pp. I--I.

\bibitem{Parthasarathi2013}
S.~H.~K. Parthasarathi, S.~Chang, J.~Cohen, N.~Morgan, and S.~Wegmann, ``{The
  Blame Game in Meeting Room ASR: An analysis of Feature Versus Model Errors in
  Noisy and Mismatched Conditions},'' in \emph{2013 IEEE International
  Conference on Acoustics, Speech and Signal Processing}, May 2013, pp.
  6758--6762.

\bibitem{Hain2007}
T.~Hain, L.~Burget, J.~Dines, G.~Garau, V.~Wan, M.~Karafiat, J.~Vepa, and
  M.~Lincoln, ``{The AMI System for the Transcription of Speech in Meetings},''
  in \emph{2007 IEEE International Conference on Acoustics, Speech and Signal
  Processing - ICASSP '07}, vol.~4, April 2007, pp. IV--357--IV--360.

\bibitem{Lincoln2005}
M.~{Lincoln}, I.~{McCowan}, J.~{Vepa}, and H.~K. {Maganti}, ``{T}he
  {M}ulti-channel {W}all {S}treet {J}ournal {A}udio {V}isual {C}orpus
  ({MC-WSJ-AV}): {S}pecification and {I}nitial {E}xperiments,'' in \emph{IEEE
  Workshop on Automatic Speech Recognition and Understanding, 2005.}, Nov 2005,
  pp. 357--362.

\bibitem{BRANDSCHAIN2010}
L.~Brandschain, D.~Graff, C.~Cieri, K.~Walker, C.~Caruso, and A.~Neely,
  ``\BIBforeignlanguage{english}{{Mixer 6}},'' in
  \emph{\BIBforeignlanguage{english}{Proceedings of the Seventh International
  Conference on Language Resources and Evaluation (LREC'10)}}, N.~C.~C. Chair),
  K.~Choukri, B.~Maegaard, J.~Mariani, J.~Odijk, S.~Piperidis, M.~Rosner, and
  D.~Tapias, Eds.\hskip 1em plus 0.5em minus 0.4em\relax Valletta, Malta:
  European Language Resources Association (ELRA), may 2010.

\bibitem{Ko2017}
T.~Ko, V.~Peddinti, D.~Povey, M.~L. Seltzer, and S.~Khudanpur, ``{A study on
  Data Augmentation of Reverberant Speech for Robust Speech Recognition},'' in
  \emph{2017 IEEE International Conference on Acoustics, Speech and Signal
  Processing (ICASSP)}, March 2017, pp. 5220--5224.

\bibitem{Barker2013}
J.~Barker, E.~Vincent, N.~Ma, H.~Christensen, and P.~Green, ``The {PASCAL}
  ``{CH}i{ME}'' {S}peech {S}eparation and {R}ecognition {C}hallenge,''
  \emph{Computer Speech \& Language}, vol.~27, pp. 621--633, 05 2013.

\bibitem{Vincent2013}
E.~Vincent, J.~Barker, S.~Watanabe, J.~L. Roux, F.~Nesta, and M.~Matassoni,
  ``The second {CH}i{ME} {S}peech {S}eparation and {R}ecognition {C}hallenge:
  {A}n overview of {C}hallenge {S}ystems and {O}utcomes,'' in \emph{2013 IEEE
  Workshop on Automatic Speech Recognition and Understanding}, Dec 2013, pp.
  162--167.

\bibitem{Barker2015}
J.~Barker, R.~Marxer, E.~Vincent, and S.~Watanabe, ``{T}he third ``{CH}i{ME}''
  {S}peech {S}eparation and {R}ecognition {C}hallenge: {D}ataset, {T}ask and
  {B}aselines,'' in \emph{2015 IEEE Workshop on Automatic Speech Recognition
  and Understanding (ASRU)}, Dec 2015, pp. 504--511.

\bibitem{Barker2018}
\BIBentryALTinterwordspacing
J.~Barker, S.~Watanabe, E.~Vincent, and J.~Trmal, ``{T}he fifth ``{CH}i{ME}''
  {S}peech {S}eparation and {R}ecognition {C}hallenge: {D}ataset, {T}ask and
  {B}aselines,'' Tech. Rep., 2018. [Online]. Available:
  \url{https://arxiv.org/pdf/1803.10609.pdf}
\BIBentrySTDinterwordspacing

\bibitem{Kinoshita2013}
K.~Kinoshita, M.~Delcroix, T.~Yoshioka, T.~Nakatani, A.~Sehr, W.~Kellermann,
  and R.~Maas, ``{T}he {REVERB} {C}hallenge: {A} {C}ommon {E}valuation
  {F}ramework for {D}ereverberation and {R}ecognition of {R}everberant
  {S}peech,'' in \emph{2013 IEEE Workshop on Applications of Signal Processing
  to Audio and Acoustics}, Oct 2013, pp. 1--4.

\bibitem{Kinoshita2016}
K.~Kinoshita, M.~Delcroix, S.~Gannot, E.~A. P.~Habets, R.~Haeb-Umbach,
  W.~Kellermann, V.~Leutnant, R.~Maas, T.~Nakatani, B.~Raj, A.~Sehr, and
  T.~Yoshioka, ``{A} {S}ummary of the {REVERB} {C}hallenge: {S}tate-of-the-art
  and {R}emaining {C}hallenges in {R}everberant {S}peech {P}rocessing
  {R}esearch,'' \emph{EURASIP Journal on Advances in Signal Processing}, vol.
  2016, no.~1, p.~7, Jan 2016.

\bibitem{Ravanelli2017}
\BIBentryALTinterwordspacing
M.~Ravanelli, P.~Svaizer, and M.~Omologo, ``{R}ealistic {M}ulti-microphone
  {D}ata {S}imulation for {D}istant {S}peech {R}ecognition,'' November 2017.
  [Online]. Available: \url{arXiv:1711.09470v1}
\BIBentrySTDinterwordspacing

\bibitem{Jeub2009}
M.~Jeub, M.~Schafer, and P.~Vary, ``{A} {B}inaural {R}oom {I}mpulse {R}esponse
  {D}atabase for the {E}valuation of {D}ereverberation {A}lgorithms,'' in
  \emph{2009 16th International Conference on Digital Signal Processing}, July
  2009, pp. 1--5.

\bibitem{Eaton2015e}
\BIBentryALTinterwordspacing
J.~Eaton, N.~D. Gaubitch, A.~H. Moore, and P.~A. Naylor, ``{Acoustic
  Characterization of Environments (ACE) Challenge Results Technical Report},''
  Imperial College London, Tech. Rep., 2017. [Online]. Available:
  \url{https://arxiv.org/pdf/1606.03365.pdf}
\BIBentrySTDinterwordspacing

\bibitem{Eaton2016}
------, ``{Estimation of Room Acoustic Parameters: The ACE Challenge},''
  \emph{IEEE/ACM Trans. Audio, Speech and Lang. Proc.}, vol.~24, no.~10, pp.
  1681--1693, Oct. 2016.

\bibitem{Nakamura2000}
S.~Nakamura, K.~Hiyane, F.~Asano, T.~Nishiura, and T.~Yamada, ``{A}coustical
  {S}ound {D}atabase in {R}eal {E}nvironments for {S}ound {S}cene
  {U}nderstanding and {H}ands-free {S}peech {R}ecognition.'' in \emph{LREC},
  2000.

\bibitem{Ravanelli2015}
M.~Ravanelli, L.~Cristoforetti, R.~Gretter, M.~Pellin, A.~Sosi, and M.~Omologo,
  ``{T}he {DIRHA}-{ENGLISH} {C}orpus and {R}elated {T}asks for {D}istant-speech
  {R}ecognition in {D}omestic {E}nvironments,'' in \emph{2015 IEEE Workshop on
  Automatic Speech Recognition and Understanding (ASRU)}, Dec 2015, pp.
  275--282.

\bibitem{Bertin2016}
N.~Bertin, E.~Camberlein, E.~Vincent, R.~Lebarbenchon, S.~Peillon,
  {\'{E}}.~Lamande, S.~Sivasankaran, F.~Bimbot, I.~Illina, A.~Tom, S.~Fleury,
  and {\'{E}}.~Jamet, ``{A} {F}rench {C}orpus for {D}istant-microphone {S}peech
  {P}rocessing in {R}eal {H}omes,'' in \emph{Proceedings of the Annual
  Conference of the International Speech Communication Association,
  INTERSPEECH, San Francisco, United States}, 09 2016, pp. 2781--2785.

\bibitem{vacher2014}
\BIBentryALTinterwordspacing
M.~Vacher, B.~Lecouteux, P.~Chahuara, F.~Portet, B.~Meillon, and N.~Bonnefond,
  ``{The Sweet-Home Speech and Multimodal Corpus for Home Automation
  Interaction},'' in \emph{{The 9th edition of the Language Resources and
  Evaluation Conference (LREC)}}, Reykjavik, Iceland, 2014, pp. 4499--4506.
  [Online]. Available: \url{http://hal.archives-ouvertes.fr/hal-00953006}
\BIBentrySTDinterwordspacing

\bibitem{Ravanelli2014}
M.~Ravanelli and M.~Omologo, ``{O}n the {S}election of the {I}mpulse
  {R}esponses for {D}istant-speech {R}ecognition based on {C}ontaminated
  {S}peech {T}raining,'' in \emph{Proceedings of the Annual Conference of the
  International Speech Communication Association, INTERSPEECH}, 09 2014, p.~4.

\bibitem{Schroeder:1979}
M.~R. Schroeder, ``{Integrated-impulse Method for Measuring Sound Decay without
  using Impulses},'' \emph{The Journal of the Acoustical Society of America},
  vol.~66, pp. 497--500, 1979.

\bibitem{Dunn1993}
C.~Dunn and M.~J. Hawksford, ``{Distortion Immunity of MLS-Derived Impulse
  Response Measurements},'' \emph{JOURNAL OF THE AUDIO ENGINEERING SOCIETY},
  vol.~41, no.~5, pp. 314--335, 1993.

\bibitem{Aoshima1981}
N.~Aoshima, ``{C}omputer-generated {P}ulse {S}ignal applied for {S}ound
  {M}easurement,'' \emph{The Journal of the Acoustical Society of America},
  vol.~69, no.~5, pp. 1484--1488, 1981.

\bibitem{Farina2000}
A.~Farina, ``{Simultaneous Measurement of Impulse Response and Distortion with
  a Swept-sine Technique},'' in \emph{Audio Engineering Society Convention
  108}, Feb 2000.

\bibitem{Stan:2002}
G.~B. Stan, J.~J. Embrechts, and D.~Archambeau, ``{Comparison of different
  Impulse Response Measurement Techniques},'' \emph{JOURNAL OF THE AUDIO
  ENGINEERING SOCIETY}, vol.~50, pp. 249--262, 2002.

\bibitem{Golomb1981}
S.~W. Golomb, \emph{{S}hift {R}egister {S}equences}.\hskip 1em plus 0.5em minus
  0.4em\relax Laguna Hills, CA, USA: Aegean Park Press, 1981.

\bibitem{Vanderkooy2012}
J.~Vanderkooy, ``{A}spects of {MLS} {M}easuring {S}ystems,'' \emph{{AES}:
  {J}ournal of the {A}udio {E}ngineering {S}ociety}, vol.~42, 01 2012.

\bibitem{Ravanelli2012}
M.~Ravanelli, A.~Sosi, P.~Svaizer, and M.~Omologo, ``{Impulse Response
  Estimation for Robust Speech Recognition in a Reverberant Environment},'' in
  \emph{2012 Proceedings of the 20th European Signal Processing Conference
  (EUSIPCO)}, Aug 2012, pp. 1668--1672.

\bibitem{Farina2007}
A.~Farina, ``{Advancements in Impulse Response Measurements by Sine Sweeps},''
  in \emph{Audio Engineering Society Convention 122}, May 2007.

\bibitem{Thomas2009}
\BIBentryALTinterwordspacing
M.~R.~P. Thomas, ``{MLS Project},'' Tech. Rep., 2009. [Online]. Available:
  \url{http://www.commsp.ee.ic.ac.uk/~mrt102/projects/mls.html}
\BIBentrySTDinterwordspacing

\bibitem{Siltanen2010}
S.~Siltanen, T.~Lokki, and L.~Savioja, ``\BIBforeignlanguage{English}{{Rays or
  Waves? Understanding the Strengths and Weaknesses of Computational Room
  Acoustics Modeling Techniques}},'' in \emph{\BIBforeignlanguage{English}{The
  International Symposium on Room Acoustics (ISRA2010), Melbourne, Australia,
  August 29-31}}, 2010.

\bibitem{Savioja2015}
L.~Savioja and U.~P. Svensson, ``{O}verview of {G}eometrical {R}oom {A}coustic
  {M}odeling {T}echniques,'' \emph{The Journal of the Acoustical Society of
  America}, vol. 138, no.~2, pp. 708--730, 2015.

\bibitem{Ciskowski1991}
R.~D. Ciskowski and C.~A. Brebbia, \emph{{B}oundary {E}lement {M}ethods in
  {A}coustics}.\hskip 1em plus 0.5em minus 0.4em\relax New York: Elsevier
  Applied Science, 1991.

\bibitem{Ihlenburgc1998}
F.~Ihlenburg, \emph{{F}inite {E}lement {A}nalysis of {A}coustic
  {S}cattering}.\hskip 1em plus 0.5em minus 0.4em\relax New York: Springer,
  c1998.

\bibitem{Allen1979}
J.~B. Allen and D.~A. Berkley, ``{Image Method for Efficiently Simulating
  Small‐room Acoustics},'' \emph{The Journal of the Acoustical Society of
  America}, vol.~65, no.~4, pp. 943--950, 1979.

\bibitem{Krokstad1968}
A.~Krokstad, S.~Strom, and S.~S{\o}rsdal, ``{C}alculating the {A}coustical
  {R}oom {R}esponse by the use of a {R}ay {T}racing {T}echnique,''
  \emph{Journal of Sound and Vibration}, vol.~8, no.~1, pp. 118--125, 1968.

\bibitem{Kim2017}
C.~Kim, A.~Misra, K.~Chin, T.~Hughes, A.~Narayanan, T.~Sainath, and
  M.~Bacchiani, ``{G}eneration of {L}arge-scale {S}imulated {U}tterances in
  {V}irtual {R}ooms to {T}rain {D}eep-{N}eural {N}etworks for {F}ar-{F}ield
  {S}peech {R}ecognition in {G}oogle {H}ome,'' in \emph{Proceedings of the
  Annual Conference of the International Speech Communication Association,
  INTERSPEECH}, 08 2017, pp. 379--383.

\bibitem{Habets2010}
\BIBentryALTinterwordspacing
E.~A.~P. Habets, ``{Room Impulse Response Generator},'' Tech. Rep., September
  2010. [Online]. Available: \url{https://github.com/ehabets/RIR-Generator}
\BIBentrySTDinterwordspacing

\bibitem{Scheibler2018}
R.~Scheibler, E.~Bezzam, and I.~Dokmanic, ``\BIBforeignlanguage{English
  (US)}{{P}yroomacoustics: {A} {P}ython {P}ackage for {A}udio {R}oom
  {S}imulation and {A}rray {P}rocessing {A}lgorithms},'' in
  \emph{\BIBforeignlanguage{English (US)}{2018 IEEE International Conference on
  Acoustics, Speech, and Signal Processing, ICASSP 2018 - Proceedings}}, vol.
  2018-April.\hskip 1em plus 0.5em minus 0.4em\relax United States: Institute
  of Electrical and Electronics Engineers Inc., 9 2018, pp. 351--355.

\bibitem{Glembek2006}
O.~Glembek, M.~Karafi{\'{a}}t, L.~Burget, and J.~{\v{C}}ernock{\'{y}},
  ``\BIBforeignlanguage{english}{{Czech Speech Recognizer for Multiple
  Environments}},'' in \emph{\BIBforeignlanguage{english}{Radioeletronika
  2006}}, 2006, pp. 1--4.

\bibitem{Grezl2014}
F.~Gr{\'{e}}zl, M.~Karafi{\'{a}}t, and K.~Vesel{\'{y}},
  ``\BIBforeignlanguage{english}{{Adaptation of Multilingual Stacked
  Bottle-neck Neural Network Structure for New Language}},'' in
  \emph{\BIBforeignlanguage{english}{Proceedings of ICASSP 2014}}.\hskip 1em
  plus 0.5em minus 0.4em\relax IEEE Signal Processing Society, 2014, pp.
  7704--7708.

\bibitem{Kuttruff2009}
H.~Kuttruff, \emph{Room Acoustics, Fifth Edition}.\hskip 1em plus 0.5em minus
  0.4em\relax CRC Press, June 2009.

\bibitem{Schroeder1965}
M.~R. Schroeder, ``New method of measuring reverberation time,'' \emph{The
  Journal of the Acoustical Society of America}, vol.~37, no.~3, pp. 409--412,
  1965.

\bibitem{Pierce89}
A.~Pierce, \emph{Acoustics: An Introduction to Its Physical Principles and
  Applications}, 06 1989, vol.~34.

\bibitem{McCowan2005}
I.~McCowan, J.~Carletta, W.~Kraaij, S.~Ashby, S.~Bourban, M.~Flynn,
  M.~Guillemot, T.~Hain, J.~Kadlec, V.~Karaiskos, M.~Kronenthal, G.~Lathoud,
  M.~Lincoln, A.~Lisowska, W.~Post, D.~Reidsma, and P.~Wellner, ``{The AMI
  Meeting Corpus},'' L.~Noldus, F.~Grieco, L.~Loijens, and P.~Zimmerman,
  Eds.\hskip 1em plus 0.5em minus 0.4em\relax Noldus Information Technology, 8
  2005, pp. 137--140.

\bibitem{Povey2011}
D.~Povey, A.~Ghoshal, G.~Boulianne, L.~Burget, O.~Glembek, N.~Goel,
  M.~Hannemann, P.~Motl{\'{i}}{\v{c}}ek, Y.~Qian, P.~Schwarz,
  J.~Silovsk{{\'y}}, G.~Stemmer, and K.~Vesel{\'{y}}, ``{The Kaldi Speech
  Recognition Toolkit},'' in \emph{IEEE 2011 Workshop on Automatic Speech
  Recognition and Understanding}.\hskip 1em plus 0.5em minus 0.4em\relax IEEE
  Signal Processing Society, Dec. 2011.

\bibitem{PeddintiIVecs2016}
V.~Peddinti, G.~Chen, D.~Povey, and S.~Khudanpur, ``{Reverberation Robust
  Acoustic Modeling using i-vectors with Time Delay Neural Networks},'' in
  \emph{INTERSPEECH}.\hskip 1em plus 0.5em minus 0.4em\relax ISCA, 2015, pp.
  2440--2444.

\end{thebibliography}
%

%

%


\begin{IEEEbiography}{Igor~Sz\"{o}ke} (M'2004) received his PhD in Information Technology in 2010 and is Assistant Professor at Brno University of Technology (BUT), Faculty of Information Technology (FIT). Since 2003, he has been with the BUT Speech@FIT research group. His research interests are machine learning and speech data mining (automatic speech recognition, spoken term detection and data augmentation). He co-founded Phonexia in 2006 and ReplayWell in 2011.
\end{IEEEbiography}

\begin{IEEEbiography}{Miroslav~Sk\'{a}cel} received his master's degree in Information Technology at Brno University of Technology (BUT), Faculty of Information Technology (FIT) in 2015. He has been a member of BUT Speech@FIT research group since 2012. He is currently working on his PhD focusing on low-resource automatic speech recognition, speech signal enhancement and deep learning.
\end{IEEEbiography}

\begin{IEEEbiography}{Ladislav~Mo\v{s}ner} received his master's degree in Information Technology at Brno University of Technology (BUT), Faculty of Information Technology (FIT) in 2017. He is a PhD student in the BUT Speech@FIT research group. His interests are multichannel processing (beamforming), noise and reverberation robustness, and speech enhancement and discriminative training in  speaker recognition.
\end{IEEEbiography}

\begin{IEEEbiography}{Jakub~Paliesek} is Master's degree student at Brno University of Technology (BUT), Faculty of Information Technology (FIT). In 2018, he successfully defended his bachelor's thesis ``Measurement of Environment Acoustics Impact on Speech Recognition Accuracy''.
\end{IEEEbiography}

\begin{IEEEbiography}{Jan ``Honza'' \v{C}ernock\'{y}} (M'2001, SM'2008) is Associate Professor and Head of the Department of Computer Graphics and Multimedia of Brno University of Technology (BUT), Faculty of Information Technology (FIT). He also serves as managing director of BUT Speech@FIT research group. His research interests include artificial intelligence, signal processing and speech data mining (speech, speaker and language recognition). He is responsible for signal and speech processing courses at FIT BUT. In 2006, he co-founded Phonexia. He is general chair of Interspeech 2021 in Brno. 
\end{IEEEbiography}






\end{document}